\documentclass[12pt]{article}

\usepackage{abstract,advdate,amssymb,amsmath,amsfonts,bbm,booktabs,eurosym,geometry,ulem,graphicx,caption,color,setspace,sectsty,comment,footmisc,caption,lipsum,mathtools,pdflscape,quoting,subfigure,array,hyperref, makecell,multirow,csquotes, fancyhdr,ragged2e, setspace, xparse}

\usepackage{indentfirst}

\usepackage[authordate, natbib]{biblatex-chicago}
\DeclareFieldFormat[article]{title}{\mkbibquote{#1}} 
\DeclareFieldFormat[thesis]{title}{\mkbibemph{#1}} 
\usepackage{newfloat}
\usepackage{hyperref}

\DeclareFloatingEnvironment[fileext=frm,placement={!ht},name=Figure]{figureappendix}

\AtBeginEnvironment{tabular}{\singlespacing}

\def\signed #1{{\leavevmode\unskip\nobreak\hfil\penalty50\hskip2em
  \hbox{}\nobreak\hfil(#1)%
  \parfillskip=0pt \finalhyphendemerits=0 \endgraf}}

\newsavebox\mybox

\makeatletter
\DeclareCiteCommand{\citetalias}
  {\usebibmacro{prenote}}
  {\usebibmacro{citeindex}%
   \printtext[bibhyperref]{\@citealias{\thefield{entrykey}}}}
  {\multicitedelim}
  {\usebibmacro{postnote}}

\DeclareCiteCommand{\citepalias}[\mkbibparens]
  {\usebibmacro{prenote}}
  {\usebibmacro{citeindex}%
   \printtext[bibhyperref]{\@citealias{\thefield{entrykey}}}}
  {\multicitedelim}
  {\usebibmacro{postnote}}
\makeatother

\defcitealias{ncrp1983}{United States Department of Justice, NCRP (1983)}
\defcitealias{ncrp1984}{United States Department of Justice, NCRP (1984)}
\defcitealias{ncrp1985}{United States Department of Justice, NCRP (1985)}
\defcitealias{ncrp1986}{United States Department of Justice, NCRP (1986)}
\defcitealias{ncrp1987}{United States Department of Justice, NCRP (1987)}
\defcitealias{ncrp1988}{United States Department of Justice, NCRP (1988)}
\defcitealias{ncrp1989}{United States Department of Justice, NCRP (1989)}
\defcitealias{ncrp1990}{United States Department of Justice, NCRP (1990)}
\defcitealias{ncrp1991}{United States Department of Justice, NCRP (1991)}
\defcitealias{ncrp1992}{United States Department of Justice, NCRP (1992)}
\defcitealias{ncrp1993}{United States Department of Justice, NCRP (1993)}
\defcitealias{ncrp1994}{United States Department of Justice, NCRP (1994)}
\defcitealias{ncrp1995}{United States Department of Justice, NCRP (1995)}
\defcitealias{ncrp1996}{United States Department of Justice, NCRP (1996)}
\defcitealias{ncrp1997}{United States Department of Justice, NCRP (1997)}
\defcitealias{ncrp1998}{United States Department of Justice, NCRP (1998)}
\defcitealias{ncrp1999}{United States Department of Justice, NCRP (1999)}
\defcitealias{ncrp2000}{United States Department of Justice, NCRP (2000--2016)}
\defcitealias{ncrp1983-2016}{United States Department of Justice, NCRP (1983--2016)}
\defcitealias{scps1990-2009}{United States Department of Justice, SCPS (1990--2009)}

\newcolumntype{L}[1]{>{\raggedright\let\newline\\arraybackslash\hspace{0pt}}m{#1}}
\newcolumntype{C}[1]{>{\centering\let\newline\\arraybackslash\hspace{0pt}}m{#1}}
\newcolumntype{R}[1]{>{\raggedleft\let\newline\\arraybackslash\hspace{0pt}}m{#1}}

\hypersetup{
    colorlinks=true,
    linkcolor=blue,
    filecolor=magenta,      
    urlcolor=blue,
    citecolor=blue
}

\fancypagestyle{plain}{\fancyhf{}}

\begin{document}

\begin{titlepage}

\title{\vspace{0cm} Prosecutor Politics:\\The Impact of Election Cycles on Criminal Sentencing in the Era of Rising Incarceration

\vspace{0in} \vspace{-1em} }
\author{Chika O. Okafor\thanks{Okafor: Northwestern University Pritzker School of Law; Department of Economics (by courtesy); Kellogg School of Management (by courtesy); Faculty Fellow, Institute for Policy Research. Email: chika.okafor@law.northwestern.edu.}\\Northwestern University \\
}
\date{\today \vspace{-2em}} 
\maketitle
\begin{center}
\href{https://doi.org/10.1016/j.jpubeco.2026.105777}{CLICK HERE FOR PUBLICATION VERSION\\(\textit{JOURNAL OF PUBLIC ECONOMICS})}
\rule{0.85\textwidth}{0.4pt}\\[0.4em]
\begin{minipage}{0.85\textwidth}\small\singlespacing\noindent
Published as: Chika O. Okafor, ``Prosecutor Politics: Social Networks as a Mechanism for Discrimination,''
\emph{Journal of Public Economics} 262 (2026), 105777,
\href{https://doi.org/10.1016/j.jpubeco.2026.105777}{https://doi.org/10.1016/j.jpubeco.2026.105777}.
\copyright{} 2026 by Elsevier. All rights reserved.
Please cite the published version. Posted under a CC BY-NC-ND 4.0 license.\\
\end{minipage}\\[0.4em]
\rule{0.85\textwidth}{0.4pt}
\end{center}

\begin{abstract}
\noindent 

Exploiting variation in the timing of district attorney elections across nearly 40 states during the steepest rise in U.S. incarceration (roughly 1986--2006), this paper shows that election years increased per capita admissions and months sentenced to state prisons, with admissions rising 3 percent---approximately 63,000 additional prison admissions over the period. These effects were concentrated in the South, stronger in Republican counties, and declined as public opinion softened, largely disappearing after the era ended. Results suggest sentencing outcomes respond to voter preferences, implying that the politics underlying rising incarceration may be as important to understand as the policies themselves.

\vspace{.2in}

\bigskip
\end{abstract}
\setcounter{page}{0}
\thispagestyle{empty}
\end{titlepage}
\pagebreak \newpage

\onehalfspacing

\section{Introduction} \label{sec:introduction}
 
State and federal prisoners in the U.S. increased by over 600 percent between 1970 and 2000---from roughly 200,000 to 1.3 million.  As of 2020, the U.S. had the highest total number of incarcerated individuals in the world (with more than approximately 2.1 million people in jail or prison) and the highest rate of incarceration (at approximately 600 people per 100,000). \par

Leading scholars attest that the unprecedented rise in incarceration rates is in no small part attributable to public policy choices \citep{travis2014}.  For example, in the 1960s and 1970s, the use of incarceration was expanded in a multitude of ways, including: prison time being increasingly required for lesser offenses; sentences being lengthened for violent crimes and repeat offenders; and drug crimes becoming more severely punished.  In the 1980s, more legislative changes were adopted, such as mandatory minimums for various offenses.  And in the 1990s, Congress and more than half of states adopted “three strikes and you’re out” laws that mandated minimum sentences of 25 years or longer for relevant offenders.  Scholars attest that public policy choices represent the ``best single proximate explanation'' for the unprecedented rise in incarceration \citep{travis2014}, yet it remains unknown whether these legislative choices are the \textit{prime} cause or, instead, simply a proximate but intermediate one. \par

Public policy choices are not made in a vacuum; they can reflect political moments encouraged by the public will.   To understand how the public's preferences may have impacted rising incarceration, there is perhaps no better elected official to research than local prosecutors (frequently called district attorneys or DAs).  Many legal practitioners hold that it is DAs who guide who gets incarcerated and for how long---more so than any other actor in the criminal legal system.  The decision on whether to charge an individual, what charges to file, whether to drop a case, whether to offer a plea bargain, and the specific terms of the plea offer remains an unreviewable power of DAs.  Significantly, how district attorneys arrive at these [sometimes literally] life-and-death decisions remains under-explored.  Furthermore, given the importance of initial charging decisions and the extremely high prevalence of plea bargaining---a process mostly negotiated by the district attorney---many believe sentencing outcomes are significantly linked to the decisions and discretion of the respective district attorney.  In recent times, plea bargaining has been the mode of conviction for over 95 percent of convicted felons in counties, meaning individual sentences were strongly influenced by negotiations with the district attorney, and not by the outcomes of judge or jury trials.\footnote{One institutional perspective on the power of district attorneys can be found on the ACLU website: \href{https://www.aclu.org/blog/smart-justice/across-america-single-most-powerful-person-local-criminal-justice-systems}{https://www.aclu.org/blog/smart-justice/across-america-single-most-powerful-person-local-criminal-justice-systems}.}  Lastly, the vast majority of DAs are elected officials, which suggests their decisions might be influenced by politics (e.g., by considerations for re-election or for viability as a candidate for higher office later).\par

How do political incentives and pressures affect the decisions of district attorneys?  During the era of rising incarceration, the political landscape may have incentivized elected officials to enact harsher punishments.  The 1980s witnessed the expansion of the War on Drugs, with the coincident crack-cocaine epidemic\footnote{The term ``War on Drugs" was coined by the media after a press conference by Richard Nixon in 1971.  From 1980 to 1984 the federal annual budget of the FBI's drug enforcement units went from 8 million to 95 million.} and the doubling of handgun-related homicides between 1985 and 1990.  The 1990s saw public and political mobilization in favor of harsher punishments culminate in the landmark 1994 Crime Bill, the largest crime bill in U.S. history. And throughout this period, many politicians sought to be viewed as ``tough-on-crime."

Past research has found voters may weigh performance in election years more than in non-election years \citep[see, e.g.,][]{healy2014}.  As such, this paper studies the impact of DA election cycles on criminal sentencing outcomes.  I compile a new comprehensive dataset on the election cycles and offices sought for all district attorneys in office during the steepest rise in incarceration in U.S. history (roughly 1986--2006).  Exploiting variation in the timing of DA elections, I employ a dynamic differences-in-differences specification to find that being in a DA election year increases total admissions per capita to state prisons and total months sentenced per capita.  In assessing the magnitude of the findings, I estimate that the election year effects on admissions per capita rose approximately 3 percent---roughly 63,000 additional prison admissions over the time period covered by this project.  Similarly, the increase in months sentences per capita in election years is roughly 3 percent.

Various empirical checks in this paper reinforce the findings. For one, this paper provides evidence that the treatment effects in DA election years are not attributable to other relevant elections (e.g., judicial, mayoral, or sheriff). Furthermore, this paper demonstrates that the increased level of criminal sentencing during DA election years does not reflect similar cyclical changes in overall crime levels and arrest rates.  Lastly, this paper finds that the average delay between case filings and prison admissions is relatively short, meaning election year increases in the prison admissions rate can plausibly be linked to DA decision-making in anticipation of elections. \par

The data this paper leverages capture prosecutorial discretion at its most consequential margin---commitment to state prison.  However, the paper does not directly capture many of the earlier decisions DAs make---from declinations to charge reductions to jail sentences---that never produce a prison admission.  To the extent that electoral incentives also influence these unobserved margins, the estimates in this paper understate the full behavioral response of district attorneys to political incentives. 

What are the implications of these findings on understanding key drivers behind the unprecedented rise in incarceration?  First, the findings support the notion that voter preferences influence incarceration above and beyond legislative or public policy choices, since the empirical strategy controls for differences in laws across place and time (among other controls).  Second, after dispelling alternative explanations such as differences in arrest or crime rates, this paper provides evidence that political incentives have at least a short-term cyclical impact on DA behavior---and, in turn, on the punitiveness of the criminal legal system.  Lastly, I explore various mechanisms and find further evidence that sentencing outcomes are associated with public sentiment: (1) election effects are higher in the South; (2) election effects depend on county political ideology, being higher in Republican counties; (3) election effects declined in the era of rising incarceration, closely coinciding with softening public opinion on punishment as measured through the General Social Survey (a longitudinal survey administered yearly since the 1970s); and (4) election effects largely disappeared at the national level after the era of rising incarceration ended. Taken together, these findings suggest DA behavior and sentencing outcomes may respond to voter preferences---particularly surrounding preferences regarding the harshness of courts.\par

This paper makes four main contributions.  First, this paper has compiled a new comprehensive dataset on the election cycles of district attorneys.  I compile information on over 4,200 DAs---every district attorney in office between roughly 1986--2006.  Second, this paper provides causal evidence that DA election years increase the severity of certain criminal sentencing outcomes.  The empirical analysis covers approximately 40 states and spans over 20 years at the height of the rise in U.S. incarceration levels.  No other study has been found to date that compares election dynamics of district attorneys between U.S. regions and across time. Third, I explore key mechanisms behind the election year effects along spatial, political, and temporal lines---and find that district attorney election effects decline over the period 1986--2006, in tandem with U.S. public opinion softening regarding criminal punishment.  No prior studies have been found that evaluate the evolution in DA election year effects over time and compare them to trends in public opinion.

In short, district attorneys appear to respond to political incentives from the local electorate, which is important given the public's greater desire for punitiveness toward those convicted of crimes during the era of rising incarceration. Other elected officials, including some of the legislators who enacted legislative changes to the criminal legal system from the 1960s through the 1990s, may similarly have been responding to public sentiment.  If so, then the public policy choices during that era, in fact, represent an intermediate---rather than a prime---cause of the unprecedented rise in incarceration.  Consequently, collective approaches to transforming U.S. public opinion, and not simply technocratic approaches to designing public policy, may be instrumental in curbing historic levels of incarceration. \par

\textit{Related Literature.  }Research has explored more broadly the role of prosecutorial discretion in sentencing outcomes \citep{reinganum1988, bjerk2005, piehl2007}, bias \citep{didwania2022, harrington2025}, and its relevance to perpetuating sentencing disparities in light of reforms in determinate sentencing \citep{miethe1987, kessler1998, stemen2005}.  A couple of studies have explored the effects of district attorney election cycles in individual states, like Florida \citep{nadel2017} and North Carolina \citep{dyke2007}.  Yet none of these studies has employed a dynamic difference-in-differences approach, which allows more granular comparisons of election years to the years immediately preceding and following the election.  This identification strategy provides more compelling evidence for the existence (or absence) of election year effects.  Among other findings, this paper provides new insight into heterogeneity in DA election year effects over time, by geographic region, and based on the political ideology of the corresponding district.\footnote{\citet{dippel2019} has found significant heterogeneity across states in the context of judicial election cycles.} \par

Another contribution of this paper is its scope, which spans the majority of U.S. states over a 20-year time period, thereby providing greater insights on the role of election cycles on national trends in incarceration that more easily controls for idiosyncrasies within individual states.  Furthermore, the analysis on national incarceration trends cannot be performed by solely relying on the extant literature on judges, mainly because the vast majority of criminal sentences result from plea bargaining, a process heavily dependent on the influence of the district attorney.

Past research suggests that appointed district attorneys tend to have lower conviction rates than their elected counterparts, who might be more concerned with appearing competent  to their voters \citep{rasmusen2009}.  The literature also suggests that those DAs who are in office for longer terms tend to prosecute fewer cases, which might be a result of the fact that they do not have to worry about re-election quite as much as DAs who only hold office for four-year terms.  \citet{bandyopadhyay2014} compares data from North Carolina DA elections and appointed DA states to find that re-election incentives led to a 10\% increase in the proportion of cases taken to trial; the presence of a challenger led to an additional 15\% increase.  The paper found that when re-election pressures increase, DAs tend to take more cases to trial and they tend to offer fewer plea bargains, perhaps in order to boost their re-election probability.  \citet{pfaff2011} suggests that the convex increase in incarceration figures in the U.S. in the 1980s and 1990s is connected to an increase in the total number of felony filings per arrest (a discretionary power of the DA), as opposed to increases in arrest rates (decided by the police), increased time served per sentence (sometimes decided by a judge in the absence of a plea deal when the case goes to trial), and increased crime rates (often connected to much larger social, political, and economic factors).  \citet{neal2016} provides a contrasting perspective, finding that much of the growth in prison populations can be attributed to changes in sentencing policies.  \citet{krumholz2019} investigates the impact of district attorney political affiliation, finding that a Republican DA leads to an increase in prison admissions and new sentenced months per capita in the four years following their election, driven by higher incarceration of drug offenses.  In contrast, this paper focuses on the effects of public sentiment, comparing outcomes in predominantly Republican and non-Republican counties (among other things).

The sparse research on district attorneys does relate to a larger literature on how judges respond to political pressures\footnote{For example, \citet{lim2015} finds that newspaper coverage has impacted judicial sentencing for violent offenses by nonpartisan elected judges; \citet{berdejo2013} finds that Washington State judges have given 10\% longer sentences at the end of their political cycle than at the beginning; and \citet{abrams2023} finds that sentencing for felonies in North Carolina increases as elections approach.}---yet district attorneys and judges occupy distinct roles in the criminal legal system so may face different incentives.  \citet{huber2004} mentions how voters may be more likely to perceive instances of underpunishment by judges than overpunishment, which plausibly also applies to the prosecutorial discretion that district attorneys exercise. In addition, many judges themselves previously served as district attorneys, introducing another source of similarity through selection. \par

This paper proceeds as follows.  Section 2 provides institutional background on the role of the district attorney.  Section 3 describes the data collected for this study.  Section 4 explains the empirical strategy.  Section 5 describes the main results.  Section 6 explores heterogeneity and mechanisms behind the results related to public sentiment, and Section 7 concludes.

\section{Institutional Background} \label{sec:background}

\subsection{Description of the District Attorney's Role}
The district attorney is the most common term for the chief local prosecutor, who can also be referred to as a county prosecutor, a city prosecutor, a state prosecutor, a state attorney, a prosecuting attorney, a Commonwealth’s attorney, a circuit solicitor, an attorney general, or a district attorney general, depending on the state.  Table \ref{tab:prosecutor1} shows various characteristics of DAs by state. During the time period of this study,  there were approximately 2300 DA jurisdictions, each served by its own unique DA.  In most states, an individual DA jurisdiction corresponds with a particular county, though there are exceptions in which a single DA jurisdiction may span several counties.  As such, DAs can be the representative for either a single county, numerous counties, or an entire state in legal matters involving the government.  While DAs have complete authority to prosecute within their districts, the direct day-to-day tasks involved in the prosecution of felonies are performed by their assistant district attorneys (ADAs) and/or deputy district attorneys (DDAs).  Yet, DAs are often responsible for the actual hiring and promotion of ADAs and DDAs in their offices.  Furthermore, DAs decide the charging and sentencing policies for their office (within the discretion allowed by state law): just as they can decide to follow more punitive policies in their district, they can also decline to prosecute some crimes, they can narrow the range of sentencing dictated by the law for certain crimes, and they can choose rehabilitative sentences over incarceration. For example, on March 20, 2020, it was reported that Kim Foxx, the DA in Cook County, IL, decided that her office would not prosecute low-level drug offenses during the COVID-19 pandemic. In short, in their respective offices, district attorneys can set the tone of prosecution, the culture of the office, and the decision-making policies.

\begin{table*}[!p]
\footnotesize
\centering
\caption{Chief Prosecutors Who Handle Felony Cases in State Courts, 2001}
    \begin{tabular}{@{}lcllcc@{}}\toprule
\makecell{State} & \makecell{Number of\\Chief\\Prosecutors} & \makecell{Title} & \makecell{Areas of\\Jurisdiction}& \makecell{Elected}& \makecell{Appointed}\\ \midrule
Alabama$^1$ & 41 & District Attorney & Judicial circuit & X \\
Alaska & 1 & Attorney General & Entire state & & X \\
Arizona & 15 & County Attorney & County & X \\
Arkansas & 28 & Prosecuting Attorney & Judicial Circuit & X \\
California & 58 & District Attorney & \makecell{County, City/County\\govt of San Francisco} & X \\
Colorado & 22 & District Attorney & Judicial circuit & X \\
Connecticut & 13 & State's Attorney & \makecell{Judicial district\\(city and town based)} & & X \\
Delaware & 1 & Attorney General & Entire state & X\\
Dist. of Columbia & 1 & U.S. Attorney & \makecell{U.S. Attorney jurisdiction\\over adult criminal cases} & & X \\
Florida & 20 & State Attorney & Judicial circuit & X \\
Georgia & 48 & District Attorney & Judicial circuit & X \\
Hawaii & 4 & Prosecuting Attorney & County & X \\
Idaho & 44 & Prosecuting Attorney & County & X \\
Illinois & 102 & State's Attorney & County & X \\
Indiana & 90 & Prosecuting Attorney & Judicial circuit & X \\
Iowa & 99 & County Attorney & County & X \\
Kansas & 105 & County Attorney$^2$ & County & X \\
Kentucky & 56 & \makecell{Commonwealth's\\Attorney} & Judicial circuit & X \\
Louisiana & 41 & District Attorney & Judicial circuit, Orleans & X \\
Maine & 8 & District Attorney & Geographical district & X \\
Maryland & 24 & State's Attorney & County, Baltimore City & X \\
Massachusetts & 11 & District Attorney & Geographical district & X \\
Michigan & 83 & Prosecuting Attorney & County & X \\
Minnesota & 87 & County Attorney & County & X \\
Mississippi & 22 & District Attorney & Judicial circuit & X \\
Missouri & 115 & Prosecuting Attorney$^3$ & County & X \\
Montana & 55 & County Attorney & County & X \\
Nebraska & 93 & County Attorney & County & X \\
Nevada & 17 & District Attorney & County, Carson City & X \\
New Hampshire & 10 & County Attorney & County & X \\
New Jersey & 21 & County Prosecutor & County & & X \\
 \toprule
\end{tabular}
\caption*{\footnotesize 
Source: \href{https://web.archive.org/web/20220611193124/https://www.cga.ct.gov/2003/rpt/2003-R-0231.htm}{https://web.archive.org/web/20220611193124/https://www.cga.ct.gov/2003/rpt/2003-R-0231.htm}
$^1$ The 10th Judicial Circuit is divided into two prosecutorial districts, each with a district attorney.\\
$^2$ Called District Attorney in five counties.\\
$^3$ Called Circuit Attorney in the city of St. Louis.}
\label{tab:prosecutor1}
\end{table*} %

\begin{table*}[!tb]
\footnotesize
\centering

    \begin{tabular}{@{}lcllcc@{}}\toprule
\makecell{State} & \makecell{Number of\\Chief\\Prosecutors} & \makecell{Title} & \makecell{Areas of\\Jurisdiction}& \makecell{Elected}& \makecell{Appointed}\\ \midrule
New Mexico & 14 & District Attorney & Judicial district & X \\
New York & 62 & District Attorney & \makecell{County (5 boroughs\\of New York City)} & X \\
North Carolina & 39 & District Attorney & Prosecutorial district & X \\
North Dakota & 53 & State's Attorney & County & X \\
Ohio & 88 & Prosecuting Attorney & County & X \\
Oklahoma & 27 & District Attorney & Judicial district & X \\
Oregon & 36 & District Attorney & County & X \\
Pennsylvania & 67 & District Attorney & \makecell{County, City/County\\government} &  X \\
Rhode Island & 1 & Attorney General & Entire state & X \\
South Carolina & 16 & Solicitors & Judicial district & X \\
South Dakota & 66 & State's Attorney & County & X \\
Tennessee & 31 & \makecell{District Attorneys\\General} & Judicial district & X \\
Texas$^4$ & 155 & \makecell{District Attorney,\\Criminal District\\Attorney, and County\\and District Attorney} & County, judicial district & X \\
Utah & 29 & County Attorney$^5$ & County & X \\
Vermont & 14 & State's Attorney & County & X \\
Virginia & 120 & \makecell{Commonwealth's\\Attorney} & \makecell{County, 26\\independent cities} & X \\
Washington & 39 & Prosecuting Attorney & County & X \\
West Virginia & 55 & Prosecuting Attorney & County & X \\
Wisconsin & 71 & District Attorney & County$^6$ & X \\
Wyoming & 23 & \makecell{District Attorney,\\County and\\Prosecuting Attorney} & \makecell{Judicial district\\County where district\\attorney has office} & X \\
\textbf{Total} & \textbf{2,341} & & & 47 & 4 \\
 \toprule
\end{tabular}
\caption*{\footnotesize 
$^4$ Four counties (Kerr, Pecos, Reagan, and Tom Green) have 2 district attorneys with concurrent jurisdiction.
$^5$ Called District Attorney in Salt Lake County.\\
$^6$ Two counties share a district attorney.}
\label{tab:prosecutor2}
\end{table*} %

\subsection{Selection Process}

Currently, chief local prosecutors are elected in every U.S. state/area but four---Alaska, Connecticut, New Jersey, and Washington DC, where they are appointed respectively by their Attorney general, a commission, or the governor.  After being elected or appointed, DAs in most states serve four-year terms---those in Alabama, Kentucky, and Louisiana serve six-year terms, DAs in Tennessee serve eight-year terms, and DAs in New Hampshire serve two-year terms.  During the period studied here, prosecuting attorneys in Arkansas (until the 2002 election) and district attorneys in Wisconsin (until the 2008 election) also served two-year terms.  In every state except Colorado, there is no limit to the number of terms that one can hold the office of district attorney.  There is also not an equal distribution of the population that a given DA district serves---the DA’s office in Los Angeles oversees over 10 million people, while the office in Alpine County, California, has jurisdiction over just below 1000 people \citep{hessick2020}. \par

\subsection{Electoral Incentives Without Electoral Competition} \label{sec:electoralincentives}
A basic question arises from this paper’s setting: why would electoral pressures arise for district attorneys even absent direct competition?  I propose three complementary reasons why.  First, career concerns: district attorneys are concerned not only about the next election, but also about their next office.  And performance during electoral periods is inherently more salient, creating signals for parties and donors when pursuing higher office.  This channel is consistent with canonical career concern models in which current performance affects future prospects \citep{holmstrom1999}.  Second, deterrence of challengers: even if a district attorney runs unopposed in a given election, they still face the threat of future competition.  Especially in climates with higher voter appetite for punitiveness, establishing a reputation for ``toughness’’ near elections can discourage future challengers for the office.  This dynamic is consistent with evidence from courts that shows that the potential for competition---and not just actual competition itself---shifts behavior toward more punitive decisions \citep{huber2004, gordon2007, berdejo2013}.  Third, heightened visibility: election periods increase media attention on those campaigning for office, including district attorneys.  As a result, high-profile failures near elections can both damage career trajectories and encourage future challengers for the office \citep{snyder2010}.  Importantly, all three of these mechanisms operate in a context where DA elections are rarely contested, so political incentives may operate through signaling, deterrence,  and visibility rather than just head-to-head competition.  The heterogeneity results presented in Section \ref{sec:mechanisms} are consistent with these channels.  Namely, election year effects are greater in Republican counties and in the South, areas where tough-on-crime signaling may have carried more career value during this era.  Also, election year effects declined over time as public opinion shifted, reducing the benefits of punitiveness.

\section{Data} \label{sec:data}
\subsection{Sentencing Data} \label{sec:sentencingdata}

\begin{figure}[!b]
\captionsetup{justification=centering}
\caption{United States Prison Population}
\centering
\includegraphics[width=1\textwidth]{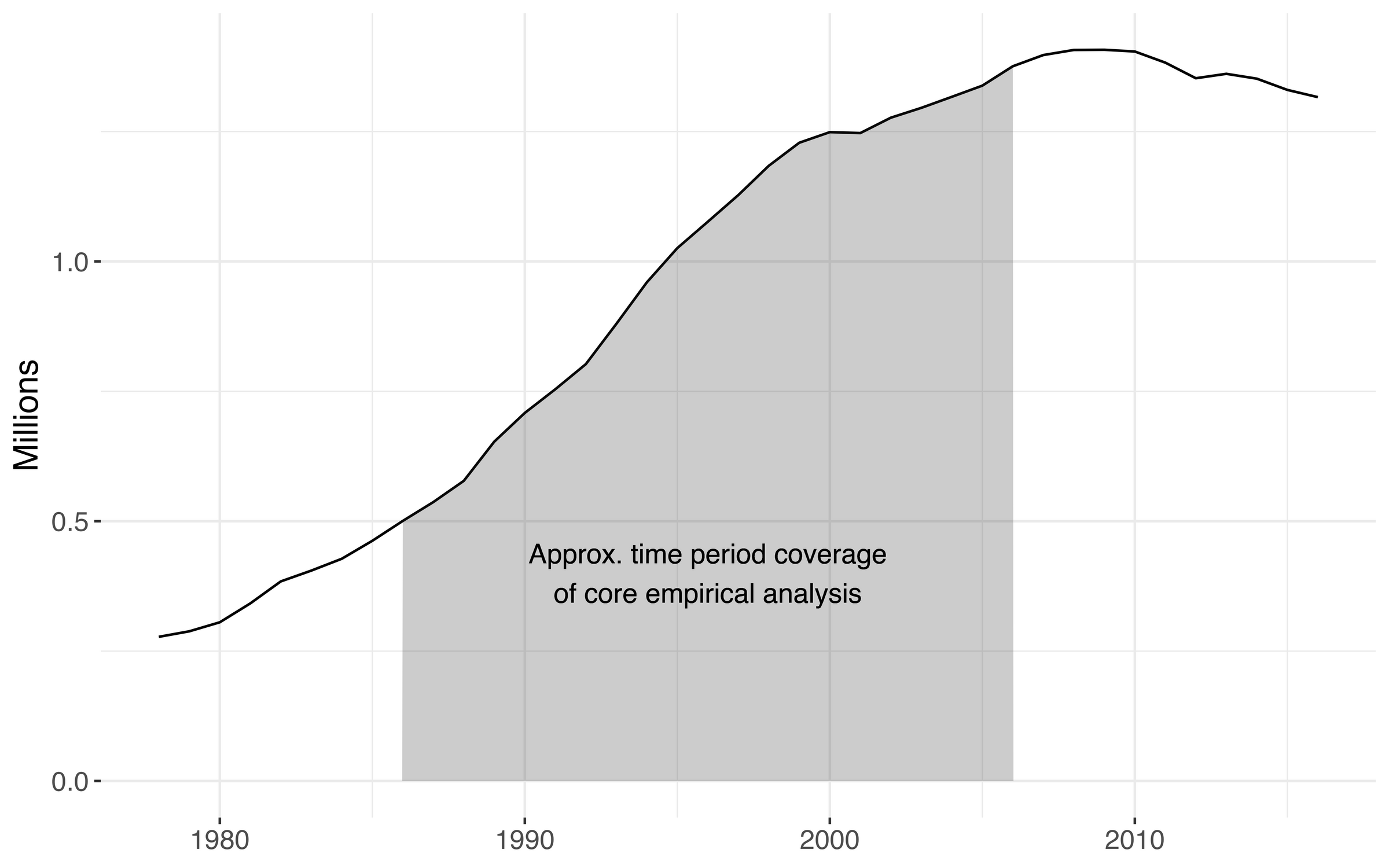} \par
\justify \footnotesize \textit{Note:} The source of the prison population data is the United States Department of Justice Office of Justice Programs (Bureau of Justice Statistics).  The core empirical analysis in this study includes entire term of all district attorneys listed in the issues of the National Directory of Prosecuting Attorneys found between 1987 and 2005.  Consequently, the coverage years indicated in the graph are approximate values, with only slight variation by district, because DAs have different start/end years for their 4-year terms based on the respective district.
\label{fig:prisonpop}
\end{figure}

\textit{National Directory of Prosecuting Attorneys---}I compiled the names of all district attorneys in the United States using the National Directory of Prosecuting Attorneys (administered by the National District Attorneys Association) for all years in which the directory could be retrieved: 1987, 1989, 1992, 1994, 1998, 1999, 2000, 2003, and 2005.\footnote{No issues of the National Directory of Prosecuting Attorneys could be found after 2005, suggesting it has been out of print for several years.}  For the core empirical analysis of this paper, I include the full term in office for the district attorneys identified through this process.  Since election years tend to fall in even-numbered years, the core empirical analysis thus spans roughly 1986--2006.  The advantages of analyzing this time period are numerous; two include: (1) the time period corresponds with the largest recorded increase in total U.S. prison population (see Figure \ref{fig:prisonpop}); and (2) the period corresponds with the greatest state participation in sentencing outcome data from the National Corrections Reporting Program (as described immediately below). \par

\textit{National Corrections Reporting Program (NCRP)---}This represents the primary source for sentencing data used in this study.  NCRP collects offender-level data on admissions and releases from state and federal prisons.  The NCRP records include offender-level information such as: BJS offense category, state offense category, total sentence, race, gender, age, prior felony status, county, and state where sentence imposed. The analysis includes the full universe of reported records on prison admissions. The National Corrections Reporting Program (NCRP) relies on data submitted voluntarily by individual states.\footnote{Reporting rates are not 100\% within state or by year (see Appendix Table C.1 of number of states included), but to bias our regressions, reporting would need to be systematically correlated with the underlying DA election cycles.  There is no evidence this is the case.  At the time of writing the manuscript, the datasets include \citetalias{ncrp1983-2016}.}  Sentenced months corresponds with the longest length of time as stated by the court that the offender could be required to serve for all offenses of which they have been convicted.\footnote{Life sentences and death penalties are coded as 1200 months, or 100 years.  The results are robust to changes in this imputation.}  All per capita sentencing outcomes included in this paper are calculated using U.S. Census estimates for the corresponding year.  The NCRP documents admissions to state prisons, which is the final stage of the criminal process. However, most cases in which district attorneys exercise discretion did not result in prison admissions.  Roughly 40 percent of state felony convictions resulted in a sentence to state prison, with the remainder largely resulting in jail or probation---and many felony arrests did not result in convictions at all.\footnote{The Bureau of Justice Statistics reports that in 2006, 460,600 felons were sentenced to prison, 321,450 were sentenced to jail, 309,730 were sent to probation, and 40,510 had other outcomes: \href{https://bjs.ojp.gov/content/pub/pdf/fssc06st.pdf}{https://bjs.ojp.gov/content/pub/pdf/fssc06st.pdf}.} The estimates in this paper therefore isolate electoral effects on the imprisonment margin of prosecutorial discretion.\par

\subsection{Election Data \& Summary Statistics} \label{sec:electiondata}

Election data was compiled from various sources.  Election information for district attorneys were compiled from states, counties, and other sources (see dataset indicated in \citet{hessick2020}).  The Database on Ideology, Money in Politics, and Elections (DIME) contains over 130 million political contributions made by individuals and organizations for local, state, and federal elections between 1979 and 2014; it also includes information on candidates and committees for state and federal elections \citep{bonica_2015data}. \par

Both the election year and election month variables described in Section \ref{sec:empiricalstrategy} are derived from the date of the general election.  Yet some incumbents face no challenger in the general election, and in heavily partisan districts, electoral pressure may subside once the primary election concludes---months before the general election.  Two features of the empirical design nonetheless limit the resulting mismeasurement.  First, primaries still fall within the calendar year of the general election.  This means electoral pressure around the primaries still falls both within the election year $t=0$ of Figure \ref{fig:alloffenses} and in the twelve-month pre-election window of Figure \ref{fig:offensesmonthly}.  Second, even if a district lacks meaningful competition in a given general election, the time period is still coded as an election year.  Although electoral incentives can operate even absent direct competition (Section \ref{sec:electoralincentives}), to the extent that such districts experience less electoral pressure, their inclusion in the treatment group would attenuate the estimates toward zero.

\begin{table}[!htbp] \centering
\caption{Summary Statistics}
\label{tab:sumstats}
\footnotesize
\textit{Panel A: Full sample} \\[2pt]
\begin{tabular}{lcccccc}
\hline\hline
 & Mean & SD & P25 & Median & P75 & N \\
\hline
Prison admissions/1{,}000 pop. & 1.74 & 3.21 & 0.64 & 1.24 & 2.25 & 41,157 \\
\quad Violent offenses$^{\dagger}$ & 0.43 & 0.80 & 0.18 & 0.32 & 0.54 & 37,950 \\
\quad Property offenses$^{\dagger}$ & 0.65 & 0.98 & 0.25 & 0.48 & 0.85 & 38,198 \\
\quad Drug offenses$^{\dagger}$ & 0.52 & 1.34 & 0.13 & 0.31 & 0.65 & 33,683 \\
Sentenced months/1{,}000 pop.$^{\dagger}$ & 146.03 & 237.59 & 48.61 & 102.24 & 192.47 & 39,255 \\
\quad Violent offenses$^{\dagger}$ & 67.30 & 130.34 & 19.35 & 44.19 & 85.28 & 36,203 \\
\quad Property offenses$^{\dagger}$ & 41.40 & 56.58 & 12.62 & 28.52 & 54.87 & 36,192 \\
\quad Drug offenses$^{\dagger}$ & 37.18 & 75.61 & 7.11 & 19.09 & 45.88 & 32,178 \\
Arrests/1{,}000 pop.$^{\ddagger}$ & 43.45 & 41.44 & 25.64 & 38.27 & 54.13 & 41,134 \\
\quad Violent offenses$^{\ddagger}$ & 5.25 & 4.76 & 2.58 & 4.49 & 6.86 & 41,134 \\
\quad Property offenses$^{\ddagger}$ & 7.05 & 7.19 & 3.39 & 5.79 & 8.96 & 41,134 \\
\quad Drug offenses$^{\ddagger}$ & 3.71 & 4.99 & 1.28 & 2.73 & 4.83 & 41,134 \\
\quad Other offenses$^{\ddagger}$ & 27.44 & 28.49 & 15.00 & 23.20 & 34.35 & 41,134 \\
Reported index crimes/1{,}000 pop.$^{\ddagger}$ & 38.91 & 26.70 & 22.36 & 34.69 & 50.69 & 41,153 \\
County population (000s) & 100.76 & 341.03 & 12.39 & 27.13 & 67.08 & 41,157 \\
Share white & 0.88 & 0.16 & 0.83 & 0.95 & 0.98 & 41,157 \\
Per-capita income (\$000s) & 21.67 & 7.75 & 16.11 & 20.79 & 25.89 & 40,349 \\
\hline\hline
\end{tabular} \\[8pt]
\textit{Panel B: Means by position in the electoral cycle} \\[2pt]
\begin{tabular}{lcccccc}
\hline\hline
 & \multicolumn{2}{c}{Admissions} & \multicolumn{2}{c}{Sent.\ months} & Arrests & \\
 & Mean & Wtd. & Mean & Wtd. & Mean & N \\
\hline
Mid-cycle & 1.71 & 2.00 & 149.0 & 162.0 & 43.8 & 10,096 \\
Year before & 1.78 & 2.03 & 146.3 & 159.3 & 43.7 & 10,403 \\
Election year & 1.86 & 2.13 & 153.8 & 165.6 & 43.5 & 10,586 \\
Year after & 1.62 & 1.90 & 134.3 & 151.6 & 42.8 & 10,072 \\
\hline\hline
\end{tabular}
\begin{minipage}{0.92\textwidth}\vspace{2pt}\footnotesize
Notes: The sample comprises county-years 1984--2008 falling within the four-year electoral cycles of district-attorney elections held 1986--2006, in districts with four-year terms and identified prosecutors (41,157 county-years; 2,059 districts). Rates are per 1{,}000 county residents. Prison admissions and sentenced months are constructed from the National Corrections Reporting Program (NCRP); arrest and reported index-crime rates are from the FBI Uniform Crime Reports; county population and demographic shares are from SEER. Per-capita income, from the Bureau of Economic Analysis, is in current (nominal) dollars. Election dates and prosecutor identities are from the author's data collection described in Section \ref{sec:data}. $^{\dagger}$Conditional on positive reported values in the category. $^{\ddagger}$UCR coverage varies by county-year, and reported crimes comprise index offenses only, so arrest totals can exceed reported crimes. Weighted means use county population, matching the weighting of subsequent regression analyses in this paper. Raw differences across positions in the election cycle reflect both electoral timing and the composition of districts observed at each position in a given year; the difference-in-differences estimates in Section \ref{sec:results} isolate the former factor. Mean arrest levels show no election-year elevation; Section \ref{sec:parallel_arrestcrime} examines this relationship more formally.
\end{minipage}
\end{table}

Table \ref{tab:sumstats} reports summary statistics of the sample used in this study.  Panel B shows that population-weighted mean admissions rise from 2.00 per 1,000 residents mid-cycle to 2.13 in election years.  This is a difference of 6.5\%, which then falls to 1.90 in the year afterward.  Similarly, the population-weighted mean sentenced months follow the same path, rising from 162 per 1,000 residents mid-cycle to 165.6 in election years, and then falling to 151.6 in the year afterward.  In contrast, the mean arrest levels, do not rise during DA election years, which is corroborated by the robustness checks in Section \ref{sec:parallel_arrestcrime}.  The election cycle trends in Panel B reflect some combination of electoral timing itself and the varying composition of districts and calendar years observed at each position in a given election cycle.  The empirical strategy in Section \ref{sec:empiricalstrategy} isolates the magnitude of the former factor.

\section{Empirical Strategy} 
\subsection{Research Design}
\label{sec:empiricalstrategy}

In this paper, I adopt a quasi-experimental research design to estimate the effects of election cycles on the decision-making of district attorneys.  County-level variation in the timing of district attorney elections resulted in substantial variation across counties that I exploit using a series of dynamic difference-in-differences specifications:%
\begin{align}
 \log (Y_{cdt}) = \gamma_d + \lambda_t + \sum_{k=-T}^{T-1} \beta_k \mathbbm{1}\{R_{ct}=k\} + \Gamma_{cdt} + \varepsilon_{cdt}
 \label{eq:dynamicdid}
\end{align}

\noindent The variable $Y_{cdt}$ corresponds to one of two outcome variables---(1) admissions per capita or (2) months sentenced per capita---evaluated for county $c$ in district $d$ in time period $t$.\footnote{Following the approach of \citet{krumholz2019}, I create per capita sentencing outcome measures aggregated from individual cases.  The usage of per capita outcomes is consistent with related research on electoral cycles in police hiring \citep{levitt2002}.  Following the approach of \citet{lim2015}, all life sentences and death penalties are coded as a 1200-month sentence.}   Time period $t$ is evaluated in either years or in months.  On the right-hand side, let $R_{ct}=t-E_c$ denote the ``relative time"---the number of periods relative to the nearest election period for the district attorney in county $c$ in time period $t$.  $T$ equals half the length of a district attorney's term in office.\footnote{The length of a district attorney term in office is determined by the state and usually equals four years (since -2 and +2 are functionally equivalent in this specification, the index of summation ends at $T-1$).  Furthermore, two states were found that had their district attorney term length change during the time period under analysis, Arkansas and Wisconsin.  Arkansas Prosecuting Attorneys were elected for 2-year terms until about 2001; Amendment 80 §20 of the Arkansas Constitution (passed in 2001) modified the Prosecuting Attorneys' term limits to 4 years. Wisconsin District Attorneys were elected for 2-year terms until about 2009 (Wis. Stat. § 978.01).}  $\gamma_d$ are district fixed effects, $\lambda_t$ are year fixed effects, and $\Gamma_{cdt}$ is a vector of controls.\footnote{The main results include controls for white share of the population and per capita income.}  The standard errors are corrected for correlation across district attorneys and over time in a given county by clustering at the district level.  \par

The coefficients of interest are $\beta_{k}$, which reflect the criminal sentencing outcomes relative to the reference time period of $X$.  In each section below, I indicate which election cycle time period corresponds with the normalized time period $X$.  For example, if I normalize such that $X=-1$, that would mean for a district attorney serving in county $c$ in time period $t$, all $\beta_{k}$ coefficients would estimate the sentencing outcomes \textit{relative to} the sentencing outcomes when $R_{ct}=-1$ (the period immediately preceding the election period).  I measure the dependent variable in logs, which yields regression estimates that can approximate the percentage difference in sentencing outcomes.\footnote{A log point estimate of $a$ approximates a $(100 \cdot a)\%$ effect when the magnitude of $a$ is small.  More formally, a log point estimate of $a$ corresponds to a multiplicative effect of $e^a$. Thus, the exact percentage effect is $(e^a - 1)\times 100\%$.}  \par

I first estimate these models using the NCRP data for which I also have data on the corresponding prosecutor from the National Directory of Prosecuting Attorneys.\footnote{The years of coverage correspond to the full term length for all prosecutors who were in office between 1987 and 2005.  The states of coverage consist of the entire United States, except for the following states/areas---Alaska, Delaware, Connecticut, New Jersey, and Washington DC.  Alaska, Connecticut, New Jersey, and Washington DC have non-elected DAs.}  The NCRP data contain offender-level information, such as BJS offense category, total sentence, county, and state where sentence imposed.  The sample for each year consists of all states that reported data that had four-year election cycles.\footnote{In the NCRP dataset, all years from 1986--2006 include data on over 44 states in the United States, with 1/3 including all 50 states.  Limiting to states with four-year election cycles yields a sample of 38 states.}  

The key identifying assumption for difference-in-differences estimation strategies is the common trends assumption, which in this case is that any differential change in outcomes in district attorney election years is the result of the district attorney election.  Perhaps the greatest challenge to this identifying assumption is the confounding effects of other relevant elections with the same timing as DA election cycles---for example, of other offices related to law enforcement (e.g., sheriff elections).  These alternative elections may have direct (and indirect) effects on sentencing outcomes, which are explored in Section \ref{sec: robustness}.

\subsection{Case Processing Timelines and Within-Year Effects} \label{sec:caseprocessing}

This paper’s empirical design raises a key question: since criminal cases take time to process, can election year effects on prison admissions be attributable to prosecutorial behavior during the same election year?  Below, I provide evidence that the answer is yes.

First, data suggest that most felony cases process quickly enough to support within-year effects.  According to State Court Processing Statistics (SCPS)---a dataset that includes felony defendants in large urban districts during a time period mostly overlapping with this study---median time from arrest to adjudication was 2.79 months from 1990--2006 \citepalias{scps1990-2009}.  Furthermore, 78.2\% of cases were adjudicated within six months and 91.6\% within nine months, with the interquartile range (25th to 75th percentile) being 1.08 to 5.49 months.  This speed in processing is largely attributable to plea bargaining, which accounts for more than 95\% of felony convictions.

Second, even accounting for the additional time between adjudication and actual prison admissions, the timeline supports within-year effects.  The median lag from arrest to adjudication is 2--3 months, with less than 1--2 months afterward to admission.  This means that arrests that occur in the November through August prior to an election would typically result in prison admissions by election day in November.  This timeline is consistent with the monthly analysis in Figure \ref{fig:offensesmonthly}, which shows elevated levels of criminal sentencing beginning early in election years.

Third, prosecutors can greatly impact the timing of cases.  District attorneys can accelerate the resolution of cases in many ways: (1) filing more severe charges that encourage defendants to plead quickly; (2) changing terms in plea offers to encourage acceptance by threatening longer sentences at trial; and (3) prioritizing processing certain categories of cases over others.  Conversely, DAs can slow the resolution of cases (e.g., by postponing charging decisions until after elections).  The discretion DAs have over timing of cases suggests that election year patterns can plausibly reflect intentional rather than incidental timing.

\section{Results} \label{sec:results}

\subsection{Election Effects on Criminal Sentencing -- Outcomes by Year} \label{sec:annual}

\begin{figure}[!tb]
\captionsetup{justification=centering}
\caption{Criminal Sentencing Outcomes,\\Relative to District Attorney Election Year}
\centering
\includegraphics[width=1\textwidth]{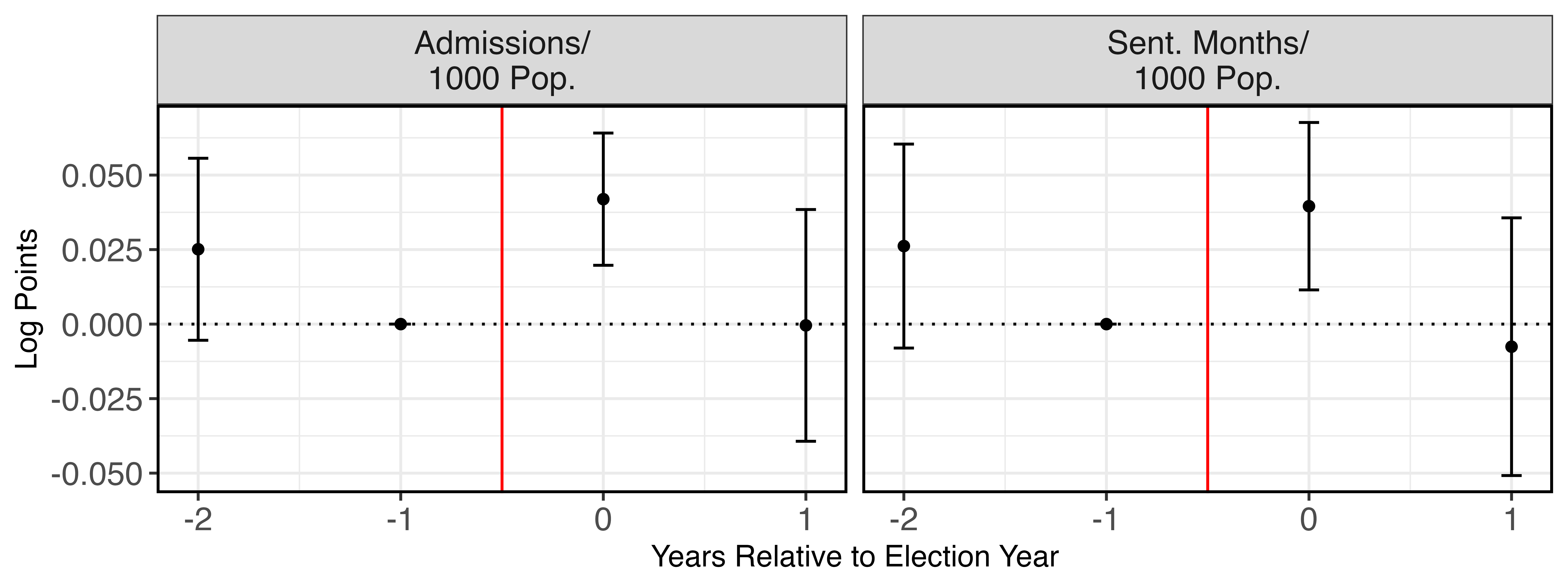} \par
\justify \footnotesize \textit{Notes:} The graph depicts dynamic difference-in-differences model estimates using Equation \ref{eq:dynamicdid} (weighted to adjust for differences in sampling probabilities across districts and time) for counties with 4-year election cycles.  Estimates are at the county level by year.  The standard errors are corrected for correlation across district attorneys and over time in a given county by clustering at the district level.  Estimates include district- and year- fixed effects.  Vertical lines mark two standard errors.  Estimates are calculated relative to the year immediately preceding the election year, which is normalized to 0.  The corresponding regression table is in the Appendix Table C.2.

\label{fig:alloffenses}
\end{figure}

Figure \ref{fig:alloffenses} graphically depicts estimates of the effects of election cycles on two separate criminal sentencing outcomes across all offenses: (1) admissions per 1000 population; and (2) total sentenced months per 1000 population.  These graphs follow the regression specification from Equation \ref{eq:dynamicdid}.  All estimates are calculated relative to the year immediately preceding the election year ($X=-1$), which is normalized to 0.  Hence, if only the estimate at time $t=0$ is positive, this can be interpreted as the election year (when $t=0$) having a relatively higher magnitude of the corresponding criminal justice outcome compared with the year prior and the year after.  All estimates are in log points.  Vertical lines mark the 95\% confidence interval. \par

The graphs provide evidence that election years have effects on the admissions rate per 1000 population and the sentenced months per 1000 population.  The estimated effects of an election year on the admissions rate are statistically significant and positive---approximately 4.3\% higher than in the year immediately preceding the election.  The estimated effects of an election year on months sentenced are also positive and statistically significant---approximately 4.0\% higher.\footnote{If these election year estimates also reflect strategic shifting of cases by district attorneys, then the treated-year estimates might incorporate both increased admissions in election years with depressed admissions in the surrounding years.}

\subsection{Election Effects on Criminal Sentencing -- Outcomes by Month} \label{sec:monthly}

\begin{figure}[!p]
\captionsetup{justification=centering}
\caption{Monthly Criminal Sentencing Outcomes by Offense Subcategory (with Adjustment for Seasonality)}
\centering
\includegraphics[width=.9\textwidth]{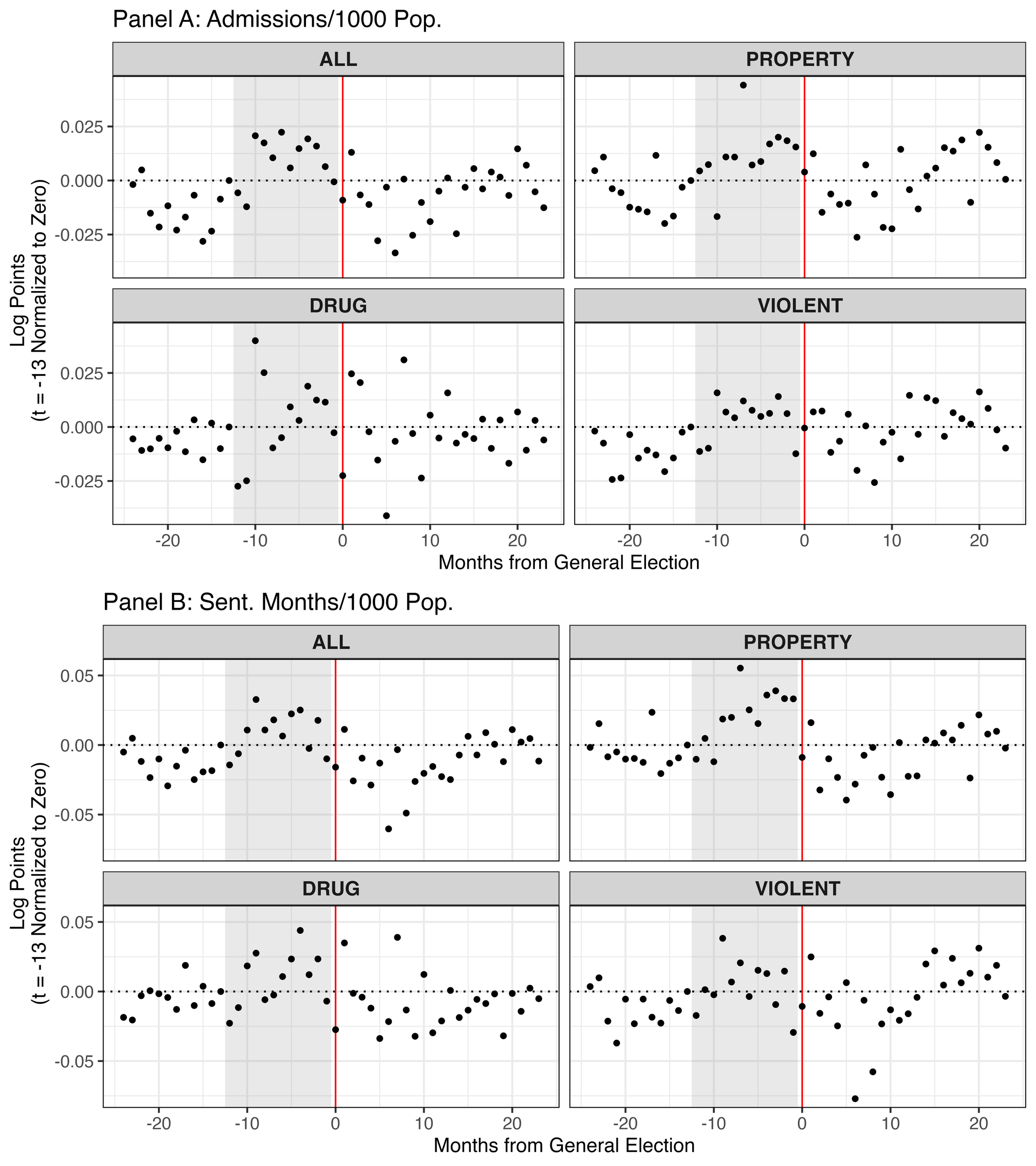} \par
\justify \footnotesize \textit{Notes:} 
The graph depicts dynamic difference-in-differences model estimates using Equation \ref{eq:dynamicdid} (population weighted to adjust for differences in sampling probabilities across districts and time). Estimates are at the county level by month. All specifications include month-of-year fixed effects estimated from non-election periods ($|t| > 12$ months from elections) to account for seasonal patterns in criminal justice outcomes. Estimates include district- and year- fixed effects. The vertical red line marks the election month (t = 0), with elections typically occurring in the first week of November. Estimates are calculated relative to t = -13 (thirteen months preceding the election month), which is normalized to 0.  The shaded region covers the twelve months preceding the general election.  The corresponding regression tables are the Appendix Tables C.11--C.14.  A plot without adjusting for seasonality is Appendix Figure C.2.
\label{fig:offensesmonthly}
\end{figure}

This subsection presents another analysis of Equation \ref{eq:dynamicdid}, this time with the time period observed in months and all coefficients measured relative to thirteen months before the election month ($X=-13$).  Figure \ref{fig:offensesmonthly} illustrates monthly estimates for total admissions (Panel A) and months sentenced (Panel B) across offense subcategories during the entire 4-year election cycle, with adjustments made for seasonality in criminal justice outcomes. \par

This analysis provides evidence that similar positive election year effects persist across offense subcategories, with the impact relative to adjacent years being most pronounced for property offenses. Though the election year estimates are positive across categories, the observed heterogeneity of election year effects across crime categories could plausibly be affected by various factors, including differences in political risk from lenient behavior and varying levels of discretionary power DAs possess for different crime categories.  First, violent cases may be the least cyclical. There is arguably greater political risk associated with being lenient on a high-profile murder over a lower-level property offense, regardless of timing in the election cycle.  Second, more severe crimes may resolve the slowest due to their complexity, pushing their outcomes past the election window even if the cases arise during the election period.  As a result, even if significant political pressure persists with violent crime, district attorneys may be more responsive in a given election cycle with their property caseload. \par

\textbf{Seasonality Adjustment.} To isolate electoral cycle effects from regular seasonal patterns in criminal justice outcomes (patterns that are clearly evident in Appendix Figure C.2 that excludes adjustments for seasonality), I implement a two-step procedure.\footnote{I control for seasonality using a two-step procedure: estimating month-of-year effects from non-election periods (more than 12 months from an election) and subtracting these from the outcome before running the main specification. This approach is appropriate given that DA elections occur in November, creating near-collinearity between calendar month and position in the election cycle.} First, I estimate month-of-year fixed effects using only observations from periods far from elections (more than 12 months before or after an election). This ensures our seasonal patterns are not contaminated by electoral pressures. Specifically, I estimate:

$$Y_{cmt} = \alpha_m + \gamma_d + \lambda_t + \varepsilon_{cmt}$$

where $Y_{cmt}$ is the outcome in county $c$ in month $m$ of year $t$, $\alpha_m$ are month-of-year fixed effects (with January as the reference month), $\gamma_d$ are district fixed effects, and $\lambda_t$ are year fixed effects. This regression is weighted by county population and estimated only on observations where $|\text{months\_to\_election}| > 12$. \par

Second, I apply the estimated monthly seasonal adjustments $\hat{\alpha}_m$ to the full dataset, creating seasonally-adjusted outcomes: $Y^{\text{adj}}_{cmt} = Y_{cmt} - \hat{\alpha}_m$. I then estimate the main specification using these adjusted outcomes, allowing one to identify electoral cycle effects net of predictable seasonal variation. \par

Figure \ref{fig:offensesmonthly} shows an upward shift in the level of criminal sentencing in the months leading up to the election.  A specification comparing the twelve months preceding the election (the shaded window in Figure \ref{fig:offensesmonthly}) with the remaining months of the four-year cycle demonstrates this point: for all offenses, admissions are 1.5 log points higher (s.e. 0.6) in the pre-election window, and months sentenced 1.8 log points higher (s.e. 0.6).  Furthermore, the period of criminal sentencing right before the election is higher than at any other period during the entire 4-year cycle for all offenses aggregated together, as well as for property offenses and drug offenses, though drug offense trends appear noisier.  As documented in Section \ref{sec:caseprocessing}, with median time from arrest to admission of just over 2--3 months, a significant majority of the prosecutorial decisions made within a year of the election would manifest as prison admissions in the months leading up to the November elections. The fact that we generally observe elevated levels of criminal sentencing beginning in January ($t=-10$) and staying elevated through the election period aligns with this timeline.  Furthermore, the lag between prosecutorial decisions and ultimate adjudication may help explain why there are not perfectly consistent declines in the admissions rates and sentenced months from $t=0$ to $t=1$.

Though both the district attorneys themselves---along with the corresponding county characteristics---may be the same across consecutive months in the same jurisdiction, individual criminal cases are generally unrelated to each other.  Thus, the fact that the dots are close together and higher in the period immediately preceding a general election suggests that there is, in fact, a higher level of criminal sentencing during this time period.  One can clearly see that all the estimates in the ten months leading up to the election month are higher magnitude than at any other ten-month period in the plot for the admissions rate outcome across all offense subcategories---as well as for the months sentenced outcome across each offense subcategory except violent offenses.  The clearest observation of this upward trend occurs in two plots: all offenses aggregated together and for property offenses taken individually.  Furthermore, the maximum monthly estimate across the entire election cycle in Panel A occurs during this pre-election period.  In Panel B, the maximum estimate similarly occurs in this pre-election period across all offense subcategories, except for drug offenses.\footnote{Appendix A develops a simple theoretical model that makes predictions about how the level of criminal sentencing will vary based on where one is in the election cycle.  The findings of this graph are consistent with the prediction of the theoretical model that the level of sentencing for district attorneys increases in the election year.} 

Lastly, the increase in criminal sentencing during the pre-election period appears more pronounced for property offenses than for violent offenses (and, to a lesser degree, for drug offenses), especially in Panel B.  This finding is consistent both with annual estimates by offense category in Appendix Figure C.1, as well as the discussion above.

\subsection{Magnitude of Election Year Effects}

Static difference-in-differences estimates (comparing the election year to all pooled non-election years) imply that election years increase prison admissions per capita by 2.8 log points (s.e. 0.8)---3 percent---with a  comparable effect on total months sentenced per capita (2.7 log points, s.e. 0.7).  The estimates from the dynamic specification comparing periods $t=0$ with $t=-1$ are larger (4.2 log points; see Figure \ref{fig:alloffenses}).  I use the static estimates to describe magnitudes below so that they draw on the full election cycle---not just adjacent years. \par

These static estimates are more conservative than the raw differences in Panel B of Table \ref{tab:sumstats}.  According to the summary statistics, population-weighted admissions levels in election years are 6.5 percent higher than in mid-cycle years.  Compared with the static estimate of approximately 3 percent, this means nearly half of the population-weighted difference can be attributed to electoral timing itself.  The remainder likely reflects differences in the composition of districts and calendar years observed across points in the election cycle. \par

These static estimates suggest that approximately 63,000 additional prison admissions are attributable to electoral timing alone across the election cycles in the sample.  This reduces to roughly 2,600 per election year.\footnote{To the extent district attorneys also shift the timing of cases in election periods, these figures can be interpreted as an upper bound on the net increase in incarceration (see also footnote 15).}  These magnitudes reflect only the cyclical component of electoral incentives---the within-district timing variation the research design isolates.  The magnitudes do not reflect level effects of electoral incentives that might cascade throughout the entire election cycle.  As such, these magnitudes do not reflect the total contribution of prosecutorial elections or broader electoral incentives to incarceration growth during this period.  Instead, they provide evidence of one mechanism through which, as Section \ref{sec:mechanisms} shows, public punitiveness during the era of rising incarceration could translate into heightened levels of criminal sentencing. 

\subsection{Robustness Checks}  \label{sec: robustness}

\subsubsection{Alternative Weighting for Two-Way Fixed Effects (TWFE) Estimator}
Recent research has highlighted that in settings with variation in treatment timing across units---such as in election cycles---the estimated coefficients in two-way fixed effects regressions can be contaminated by effects from other periods \citep{chaisemartin2020, sun2021}.  Given this potential problem in the estimation of regression coefficients in the two-way fixed effects model, I adapt an alternative estimator followin a similar logic of \citet{sun2021} (which was adapted for this setting given differences with the event-study setting they consider).\footnote{More detail on the methodology used for this alternative estimator can be found in Appendix B.}  This alternative estimator is robust to treatment effect heterogeneity.  Under this alternative two-way fixed effects estimator, estimates remain positive across all outcomes and offense subcategories; magnitudes are similar for admissions and somewhat attenuated for sentenced months (Appendix B).\par

\subsubsection{Common Trends Assumption -- Conceptual Introduction}
\label{sec:parallel}

The key identifying assumption for the difference-in-differences empirical approach used throughout this Section is common trends. Given the cyclical nature of elections, the common trends assumption in this context is different than in standard settings with an absorbing treatment (i.e., where units remain treated once initially treated).  In particular, for the common trends assumption to hold in this paper's setting, it requires that in the absence of treatment (i.e., without being in a DA election time period), the expectation of the outcome (i.e., per capita admissions rate and per capita months sentenced) follows the same evolution over time in every county \citep{chaisemartin2020}. \par

\subsubsection{Common Trends Assumption -- Empirical Analyses on Other Elections}
\label{sec:parallel_empirical}

\begin{table}[!bp] \centering
\caption{Election-Year Effects Under Alternative Specifications --- Other Elected Offices}
\label{tab:paralleltrends}
\footnotesize
\begin{tabular}{lcc}
\hline\hline
 & Admissions/1000 Pop. & Sent.\ Months/1000 Pop. \\
 & (log points) & (log points) \\
\hline
(1) Baseline & 0.028*** & 0.027*** \\
 & (0.008) & (0.007) \\
 & [40,349] & [38,483] \\ [3pt]
(2) No demographic controls & 0.028*** & 0.028*** \\
 & (0.008) & (0.007) \\
 & [41,157] & [39,255] \\ [3pt]
(3) Unweighted & 0.024*** & 0.034*** \\
 & (0.007) & (0.009) \\
 & [40,349] & [38,483] \\ [3pt]
(4) Control for synchronous judicial, & 0.007$^{\ddagger\ddagger}$ & -0.016$^{\ddagger\ddagger}$ \\
 sheriff, and mayoral elections & (0.029) & (0.022) \\
 & [34,456] & [33,644] \\ [3pt]
(5) Excl.\ synchronous judicial elections & -0.006$^{\ddagger\ddagger}$ & 0.014$^{\ddagger\ddagger}$ \\
 & (0.016) & (0.017) \\
 & [22,611] & [22,261] \\ [3pt]
(6) Excl.\ synchronous sheriff elections & 0.052** & 0.031 \\
 & (0.022) & (0.024) \\
 & [13,256] & [12,853] \\ [3pt]
(7) Excl.\ synchronous mayoral elections & 0.033*** & 0.022** \\
 & (0.013) & (0.010) \\
 & [6,293] & [6,056] \\ [3pt]
\hline\hline
\end{tabular}
\begin{minipage}{0.85\textwidth}\vspace{2pt}\scriptsize
Notes: Each cell in the Table reports estimates based on a static difference-in-differences model including an election-year indicator (weighted to adjust for differences in sampling probabilities across districts and time). The comparison group is all non-election years of the electoral cycle.  Estimates are at the county level by year.  Estimates include district- and year- fixed effects.  The standard errors are in parentheses and are corrected for correlation across district attorneys and over time in a given county by clustering at the district level.  The number of observations are in brackets. Rows (4)--(7) use the county-years for which the relevant office's election timing is observed (judicial: trial court general elections in Lim and Snyder's data, 1990--2010 for most states; sheriff: Thompson's 2003--2016 elections, backcast on each state's election cycle; mayoral: de~Benedictis-Kessner's elections for the largest city in each county).  Areas where insufficient data was found to determine synchronicity with the district attorney election cycle are omitted.  Rows (5)--(7) drop county-years in which the district attorney election year coincides with an election for the indicated office, retaining all other county-years in the same districts; row (4) instead retains all county-years in that sample and adds an indicator for such synchronous county-years.  Dynamic event study versions of these specifications appear in Appendix Figure C.5.  *~p$<$0.1, **~p$<$0.05, ***~p$<$0.01.
\par\vspace{4pt}\noindent
$^{\ddagger\ddagger}$ Rows (4) and (5) have little identifying variation and so are not informative on their own.  Specifically, where judicial election timing is observed, district attorney elections coincide with a trial court general election in nearly every jurisdiction and year.  As a result, the election year indicator used in the row (4) specification is nearly collinear with the indicator for synchronous elections, an indicator that equals one in 76 percent of the election-year county-years in that sample (83 percent by population weight).  Similarly, the row (5) specification discards the election years of the multi-court jurisdictions, which carry most of the population weight---leaving little identifying variation.  Appendix Table C.24 directly tests the concern these rows could not address, by examining jurisdictions with at least one district attorney election year that does not coincide with a trial court election (e.g., single-court counties, or benches on a different cycle) and comparing, within them, election years with and without a coinciding election.  For these jurisdictions, the analysis finds the election year increase in admissions is 2.5 log points (s.e. 1.1) in years with no coinciding judicial election, and years with a coinciding election differ by $-$0.3 log points (s.e. 1.9); for sentenced months, the corresponding estimates are 3.9 (s.e. 1.5) and $-$2.3 (s.e. 2.4).  Thus, the election year effect remains present, at the magnitude of the pooled estimate, in years in which no judicial election could account for it.
\end{minipage}
\end{table}

Perhaps the greatest threat to the common trends assumption is potential confounding effects from other elections---in other words, if there are other local and state elections systematically occurring at the same time as DA elections that meaningfully impact sentencing outcomes.  For one, judicial elections may prove particularly relevant.  Judges often exercise meaningful discretion that could impact prosecutor behavior whether within or outside of plea bargaining---negotiations that are conducted in the shadow of the trial.  The analysis below thus addresses the concern empirically rather than definitionally.  In Table \ref{tab:paralleltrends} and Appendix Table C.24, I provide empirical evidence the common trends assumption holds by testing for potential confounding effects of synchronous judicial, mayoral, and sheriff elections. 

Research has found that mayoral elections can increase local police presence \citep{levitt2002}, which may impact sentencing outcomes via changes in the arrest rate.  Judicial elections may have an impact on sentencing outcomes due to judicial discretion in sentencing for certain cases, particularly some cases where defendants are convicted via trial.  And sheriff elections may have an impact on sentencing outcomes, given the fact that sheriffs can serve as the principal police force within a county (and may thus exhibit different law enforcement behavior as elections approach).  I rely on data from \citet{lim2015more, de2018, thompson2020} for judges, mayors, and sheriffs, respectively.  The \citet{lim2015more} data consists of over 58,000 state judicial elections---at the trial, appellate and supreme court levels.  I manually map which elections by state correspond to trial court judicial elections and only use those corresponding elections in this analysis.\footnote{Trial courts are where cases are tried in the first place---as opposed to an appeals or supreme court---and represent the most plausible challenge to the common trends assumption.}  The \citet{de2018} data consists of 9,131 mayoral elections and includes data from several different sources on elections between 1950 and 2014 in cities of all sizes.  I merge the city election data with the corresponding county and take the observed election years as they are, backcasting on the city's term length for years before its first observed election.  This allows me to employ the same two-way fixed effects model I performed in the previous subsection---illustrated in Table \ref{tab:paralleltrends}---using the following two approaches: (1) controlling for the incidence of an election for another relevant office at the same time as the prosecutor elections; and (2) removing county-years in which the district attorney election year was synchronous with a local mayoral election.  For approach (2), I only include counties in which one can identify whether or not there was a synchronous mayoral election; counties without corresponding mayoral election data were not included in the analysis.  I perform a similar analysis with the sheriff election data from \citet{thompson2020}, which includes over 3,200 partisan sheriff elections, as well as the judicial data from \cite{lim2015more}.

Table \ref{tab:paralleltrends} rows (4) and (5) lack sufficient identifying variation to be informative on their own (see the table note).  Specifically, DA elections coincide with a judicial election in nearly every jurisdiction and year sampled.  This is largely because of multi-judge trial courts with staggered terms in which some judge is on the ballot in nearly every even year.  Appendix Table C.24 directly tests the common trends concern these rows could not address.  It does so by examining jurisdictions with at least one district attorney election year that does not coincide with a trial court election, comparing election years with and without a coinciding election.  In such jurisdictions, election years without a coinciding judicial election still have positive and statistically significant DA election year effects.  The election year increase in admissions is 2.5 log points (s.e. 1.1) in years with no coinciding judicial election, and years with a coinciding election differ by $-$0.3 log points (s.e. 1.9). For sentenced months, the corresponding estimates are 3.9 (s.e. 1.5) and differ by $-$2.3 (s.e. 2.4).  Thus, the election year effect persists---and near the magnitude of the pooled estimate---in years in which no judicial election could account for it.  Table \ref{tab:paralleltrends} rows (6) and (7), which exclude county-years with a coinciding sheriff or mayoral election, retain the election year effect for admissions (5.2 log points, s.e. 2.2; 3.3, s.e. 1.3).  These empirical analyses complement the conceptual discussion in the previous subsection; taken together, they address confounding due to synchronous elections for other offices.

\subsubsection{Common Trends Assumption -- Empirical Analyses on Crime Rates and Arrest Levels}
\label{sec:parallel_arrestcrime}

\begin{figure}[!b]
\captionsetup{justification=centering}
\caption{Arrest \& Crime Rates,\\Relative to District Attorney Election Year}
\centering
\includegraphics[width=1\textwidth]{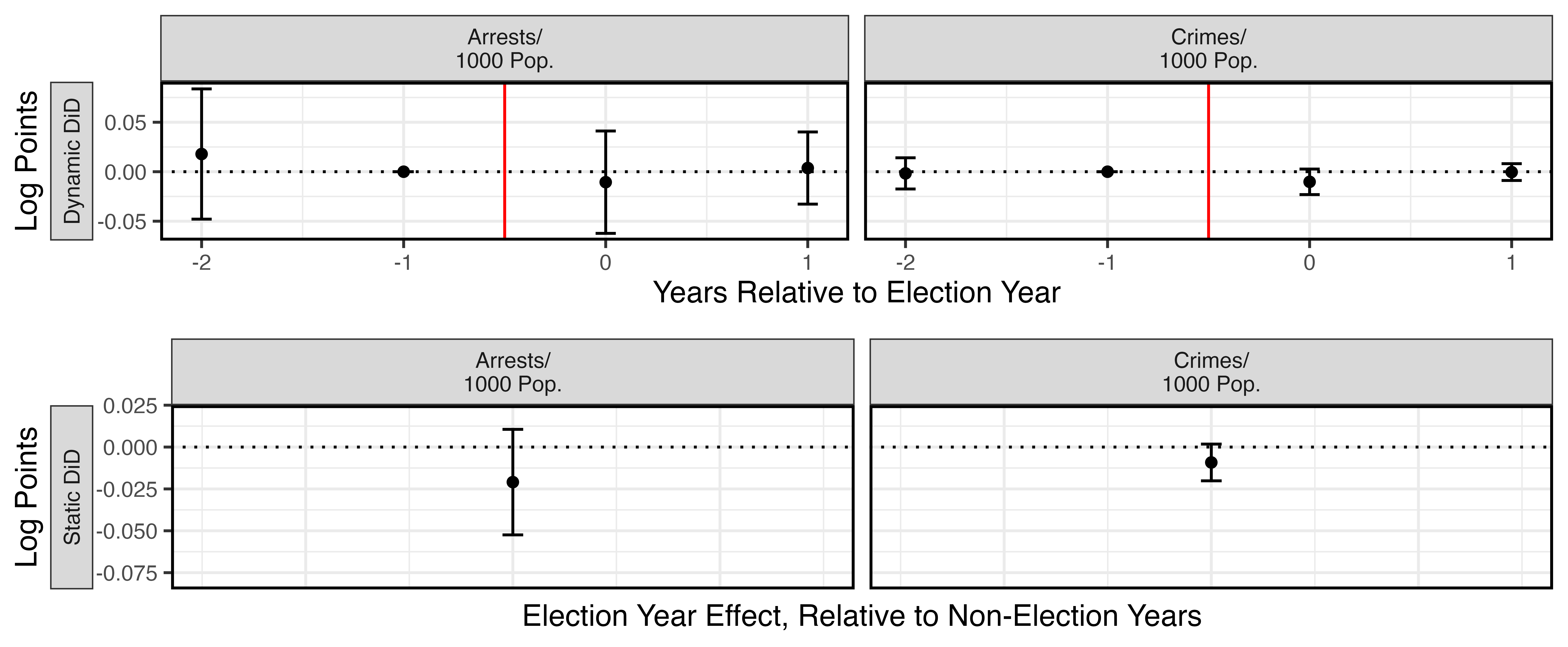} \par
\justify \footnotesize \textit{Notes:} The graphs depict dynamic or static difference-in-differences model estimates using Equation \ref{eq:dynamicdid} or Equation \ref{eq:staticdid}, respectively (weighted to adjust for differences in sampling probabilities across districts and time).  Estimates are at the county level by year.  The standard errors are corrected for correlation across district attorneys and over time in a given county by clustering at the district level.  Estimates include district- and year- fixed effects.  Vertical lines mark two standard errors.  Dynamic DiD estimates are calculated relative to the year immediately preceding the election year, which is normalized to 0.  The corresponding regression tables are the Appendix Tables C.15--C.16.

\label{fig:arrestscrimes}
\end{figure}

Another threat to the common trends assumption is that the election year increase in criminal sentencing  levels could simply reflect cyclical trends in the crime rate or volume of arrests---not anything linked to DA behavior.  Figure \ref{fig:arrestscrimes} provides evidence invalidating this threat by graphically depicting estimates of the effects of DA election cycles on per capita arrest rates and per capita crime rates,\footnote{Arrest and crime data for this analysis comes from the Uniform Crime Reporting (UCR) Program. Though there are important caveats to using the county-level statistics (see \url{https://www.icpsr.umich.edu/web/pages/NACJD/guides/ucr.html\#desc_cl}), it may represent the best available source for historical county-level arrest and crime data spanning the United States.} via the regression specification from Equation \ref{eq:dynamicdid}.  The plots show there is no statistically significant election year effect on the volume of arrests and crimes.  There appears to be no statistically significant political cycle for either the arrest rate or the crime rate.  Hence, Figure \ref{fig:arrestscrimes} does not support the notion that the main election effects on criminal sentencing outcomes can be explained by changes in the volume of arrests or crimes around elections.  In other words, this analysis undercuts a perspective that election-year increases in the admissions rate or total months sentenced might be attributable to election-year increases in arrests or crime.  Thus, it appears DA election effects are not attributable to changes in criminal behavior; this evidence suggests criminal activity does not spike in or near DA election years.  Given the aforementioned relationship between mayoral/sheriff elections, police presence, and arrest rates---and since arrest rates are not increasing in DA election years---Figure \ref{fig:arrestscrimes} suggests the relative non-synchronicity of DA elections with those of mayors and sheriffs, and/or the relatively low impact of mayoral and sheriff elections on the election effects found in this study when they are synchronous.  Appendix Table C.6 re-estimates the main specification with outcomes scaled to arrests rather than population, which shows an even greater magnitude of election year effects on admissions per arrest of approximately +3.8 log points (s.e. 1.7).  This provides further evidence that the electoral response operates through the conversion of arrests into prison admissions---the area most directly impacted by prosecutorial discretion.

\subsubsection{Other Robustness Checks}

I also performed additional robustness checks.  For example, the findings in this study are robust to specifications in which the set of controls---white share of the population and income---are not included.  The findings are also robust to county (instead of district) fixed effects.  Also, the shape of the curve depicting the relationship between the election cycle and criminal sentencing outcomes is preserved under a log transformation---log(1+x)---of the data; the utility of this transformation is that it prevents zeros from being omitted from the regression analysis.  Appendix Figures C.3--C.4 show election year estimates under a log transformation.

\section{Heterogeneity Analysis and Mechanisms on Public Sentiment} \label{sec:mechanisms}
The results thus far document that the admissions rate per capita and the rate of total months sentenced per capita increase in the lead-up to general elections during an era of rising incarceration (roughly 1986 to 2006).  In this section, I perform additional analysis to shed light on various mechanisms that underlie the relationship between DA election year sentencing outcomes and public sentiment.  The analysis uncovers several key facts, described in more detail below: (1) DA Election effects vary by region and are concentrated in South Central United States; (2) DA election effects depend on county political ideology, being higher in Republican counties; (3) DA election effects declined in the era of rising incarceration, coinciding with softening public opinion on punishment;  and (4) DA election effects disappeared on average at the national level after the era of rising incarceration ended.  Taken together, these facts provide evidence of a nexus between public sentiment and sentencing outcomes.

The approach I use in this section to estimate partial effects is the static difference-in-differences model in Equation \ref{eq:staticdid} below:%
\begin{align}
 \log(Y_{cdt}) = \alpha D_{ct} + \rho X_{ct} + \beta (D_{ct} \times X_{ct}) + \gamma_d + \lambda_t +\Gamma_{cdt} + \varepsilon_{cdt}
\label{eq:staticdid}
\end{align}

\noindent The dependent variable $Y_{cdt}$ corresponds to one of two outcome variables---(1) admissions per capita or (2) months sentenced per capita---evaluated for county $c$ in district $d$ in year $t$.  On the right-hand-side, let $D_{ct}$ denote the treatment, which is an indicator variable denoting whether county $c$ is in an election year in year $t$.  $X_{ct}$ is one of several DA or country characteristics indicated in the corresponding figures that follow.  $\gamma_d$ are district fixed effects, $\lambda_t$ are year fixed effects, and $\Gamma_{cdt}$ is a vector of controls.\footnote{Again, the main results persist with or without inclusion of controls such as white share of the population and per capita income.}  The standard errors are corrected for correlation across district attorneys and over time in a given county by clustering at the district level.  \par

\subsection{DA Election Effects Vary by Region and are Concentrated in South Central U.S.}

\begin{figure}[!tbp]
\captionsetup{justification=centering}
\caption{Election Year Effects By Region}
\centering
\includegraphics[width=1\textwidth]{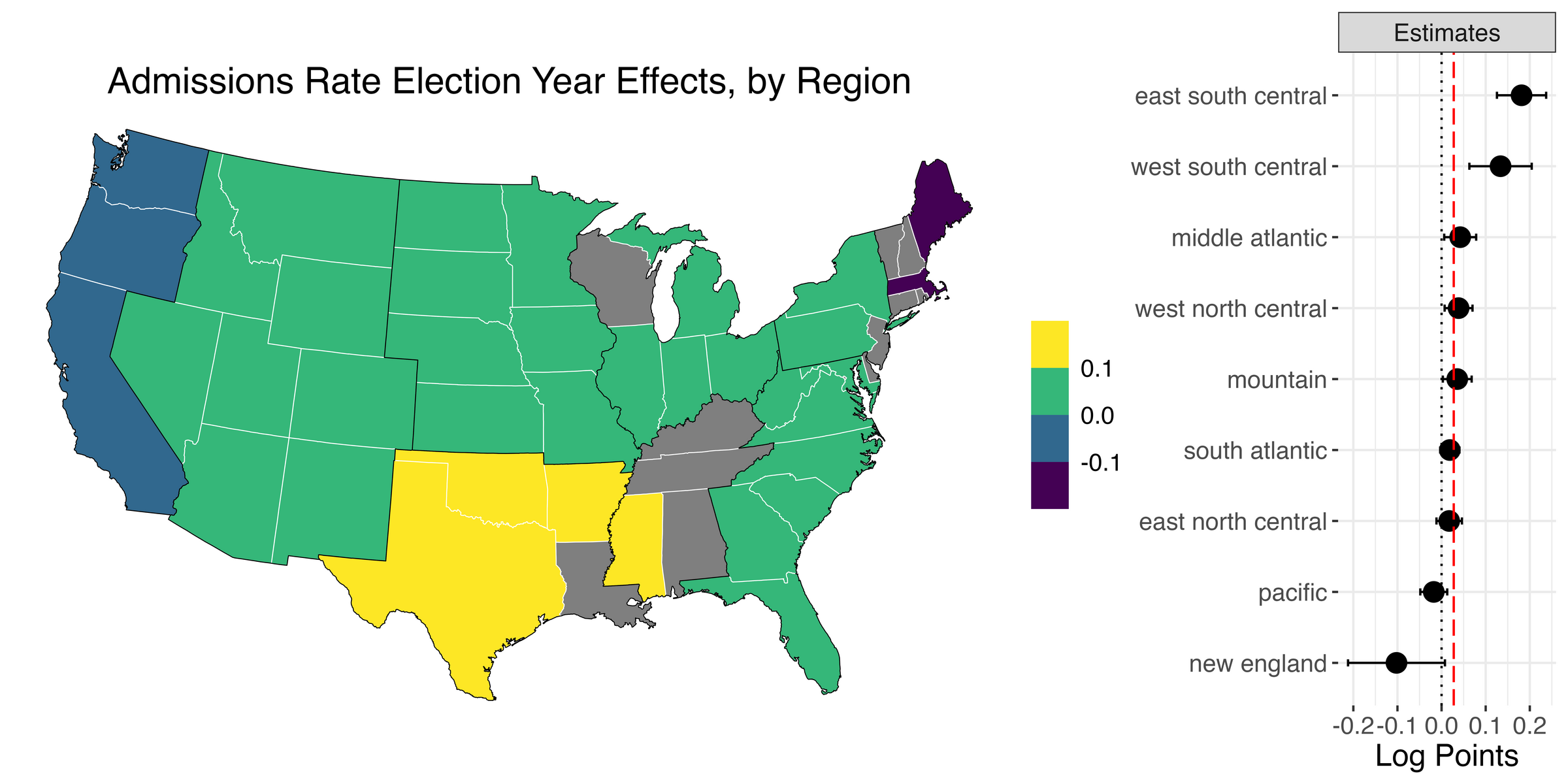} 
\medskip
\includegraphics[width=1\textwidth]{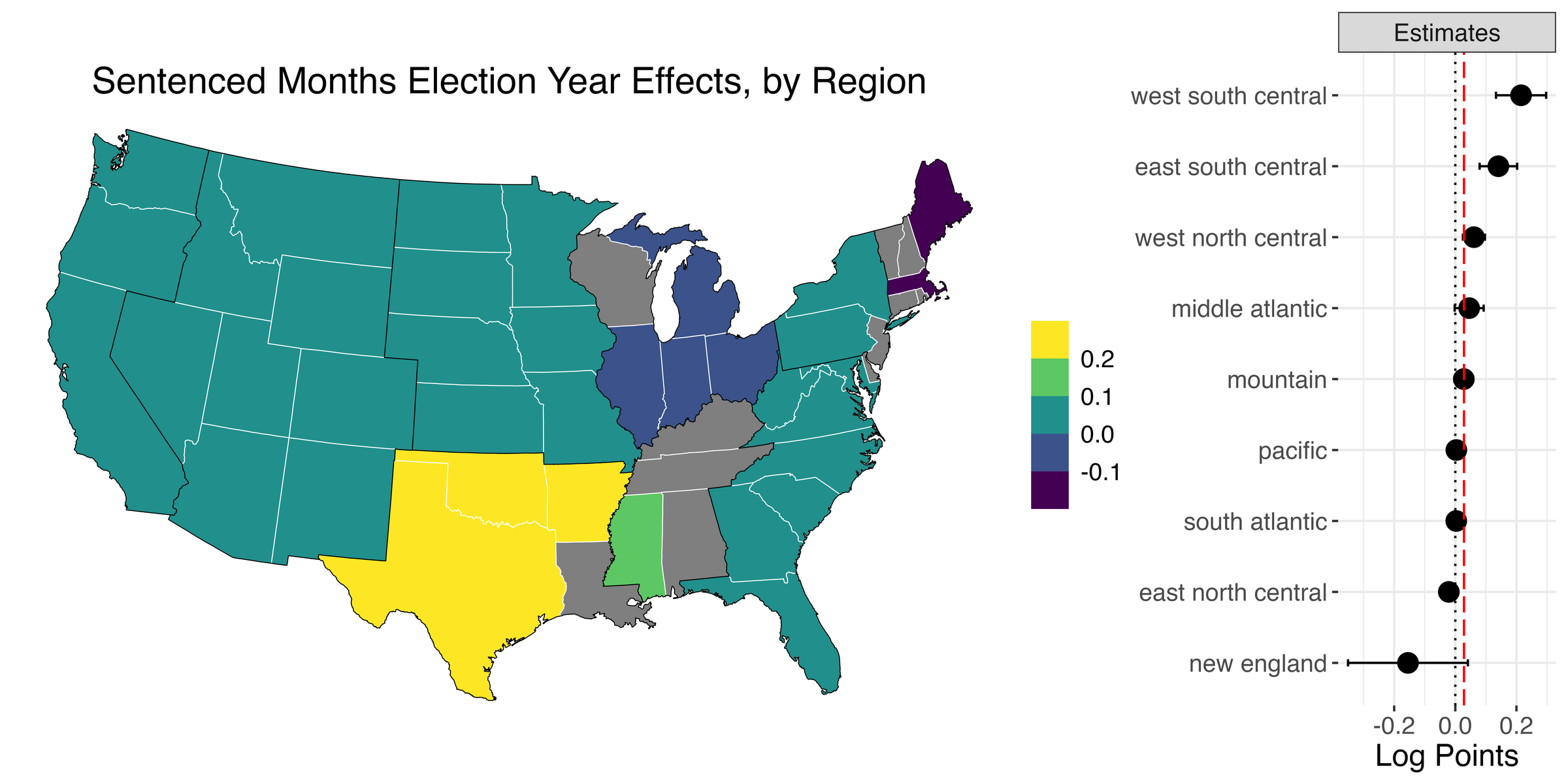}
\par
\justify \footnotesize \textit{Notes:} Map shading depicts a static difference-in-differences model (population weighted to adjust for differences in sampling probabilities across districts and time).  Estimates are at the county level by year.  The standard errors are corrected for correlation across district attorneys and over time in a given county by clustering at the district level.  Estimates include district- and year- fixed effects.  Horizontal lines mark two standard errors.  The vertical dashed line represents the main election year effect.  Heterogeneity across regions calculated via a static difference-in-difference model.  States colored gray are not included in the study sample.
\label{fig:region}
\end{figure}

Here, I examine the heterogeneity of election year effects by geographic region.  Figure \ref{fig:region} shows election year effects by region for both outcome variables, with the maps on the lefthand side shaded according to partial effects from the static difference-in-differences model (Equation \ref{eq:staticdid}).  The righthand side of the figure shows the corresponding partial effects, evaluated with the time period in years.  One can see that the largest election year effects on admissions rate is consistently in the South Central region of the U.S. (east and west).  West South Central states include Arkansas, Oklahoma, and Texas\footnote{Louisiana sentencing outcomes are not in the study sample}.  East South Central includes Mississippi\footnote{Alabama, Kentucky, and Tennessee sentencing outcomes are not in the study sample}.  The plot for the months sentenced outcome shows similar results, with the largest election year effects again for the South Central region (east and west).  

\subsection{DA Election Effects Track County Political Ideology}

\begin{figure}[!tb]
\captionsetup{justification=centering}
\caption{Association of County and Prosecutor Characteristics with Election Year Effects}
\centering
\includegraphics[width=1\textwidth]{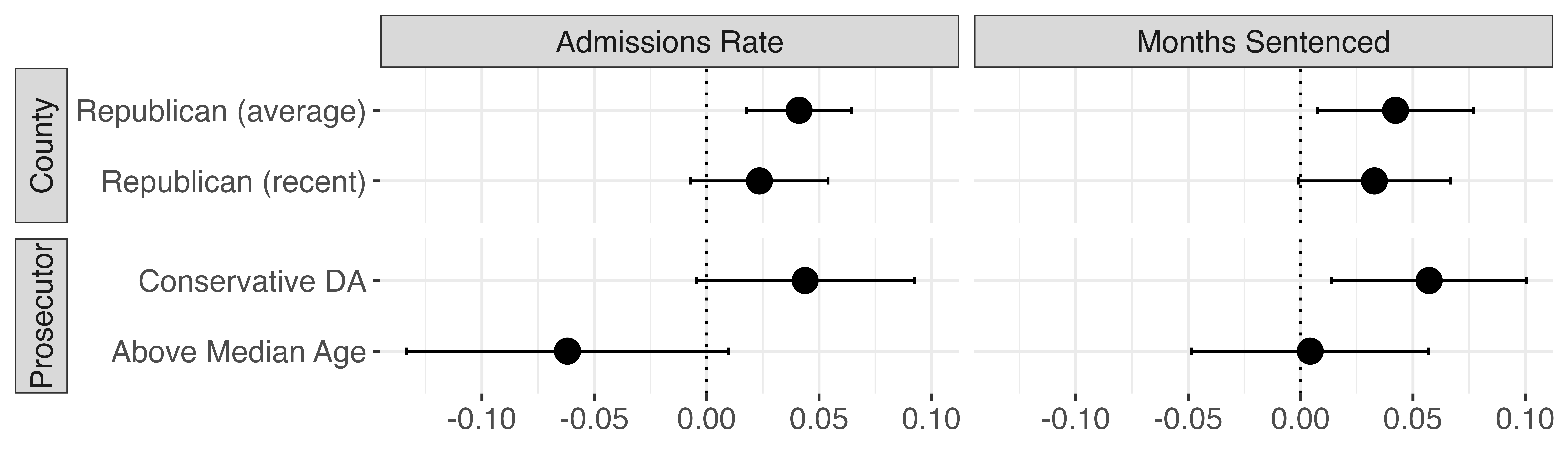} 
\par
\justify \footnotesize \textit{Notes:} The graph depicts a static difference-in-differences model (population weighted to adjust for differences in sampling probabilities across districts and time), calculated by running a pooled regression of Equation \ref{eq:staticdid}.  All specifications include main effects for each characteristic ($\rho X_{ct}$ in Equation \ref{eq:staticdid}) in addition to the interaction with the election year indicator; plotted coefficients represent the interaction terms ($\beta$).  ``Above Median Age'' is defined as college graduation year + 21 years, compared with the sample median.  Estimates are at the county level by year.  The standard errors are corrected for correlation across district attorneys and over time in a given county by clustering at the district level.  Estimates include district- and year- fixed effects.  Horizontal lines mark two standard errors.  ``Republican (recent)" corresponds with counties that voted for a Republican Presidential candidate in the most recent election.  ``Republican (average)" corresponds with counties that voted for a Republican Presidential Candidate in more than 6 out of 10 elections from 1980 to 2016.  
\label{fig:countyelectiontraits}
\end{figure}

Figure \ref{fig:countyelectiontraits} shows the association of county and prosecutor traits with DA election year effects on criminal sentencing outcomes.  The political ideology of the district has a statistically significant association with the election year effects on admissions and months sentenced, while the political ideology of the district attorneys themselves is weakly positive for admissions and statistically significant for months sentenced.\footnote{A district attorney is coded as conservative if the winner of the district's most recent DA election ran as a Republican; where party is unavailable, coding is done by linking the district attorney names with donors with the Database on Ideology, Money in Politics, and Elections (DIME) \citep{bonica_2015data}.  Individuals with an ideology score below 0 were coded as liberal, while those with a score above 0 were coded as conservative.  The coverage of the DIME database does not include all district attorneys; hence, Figure \ref{fig:countyelectiontraits} includes analysis on DAs who were successfully matched.}  Counties that voted Republican in more than 6 out of 10 elections from 1980 to 2016 have a statistically significant and positive association with election year effects for the admissions rate and months sentenced.  By contrast, counties whose most recent presidential vote favored the Republican candidate show a weakly positive association across both outcomes, with a lower magnitude than the average county voting pattern.  One interpretation of these findings is that durable local political ideology---rather than the outcome of any single election---shapes election year effects on sentencing.  This suggests DAs may respond to stable political preferences of their electorate, consistent with a mechanism for DA behavior tied to political incentives (and with the theoretical model from Appendix Equation A.1).\footnote{There may be other plausible interpretations; for example, an uptick in local crime could raise both admissions and Republican voting.  That the association is, if anything, stronger for the long-run measure than for the most recent election is not what one would expect if it were driven by transient local shocks.}

\begin{figure}[!tb]
\captionsetup{justification=centering}
\caption{Association of County Characteristics with Election Year Effects}
\centering
\includegraphics[width=1\textwidth]{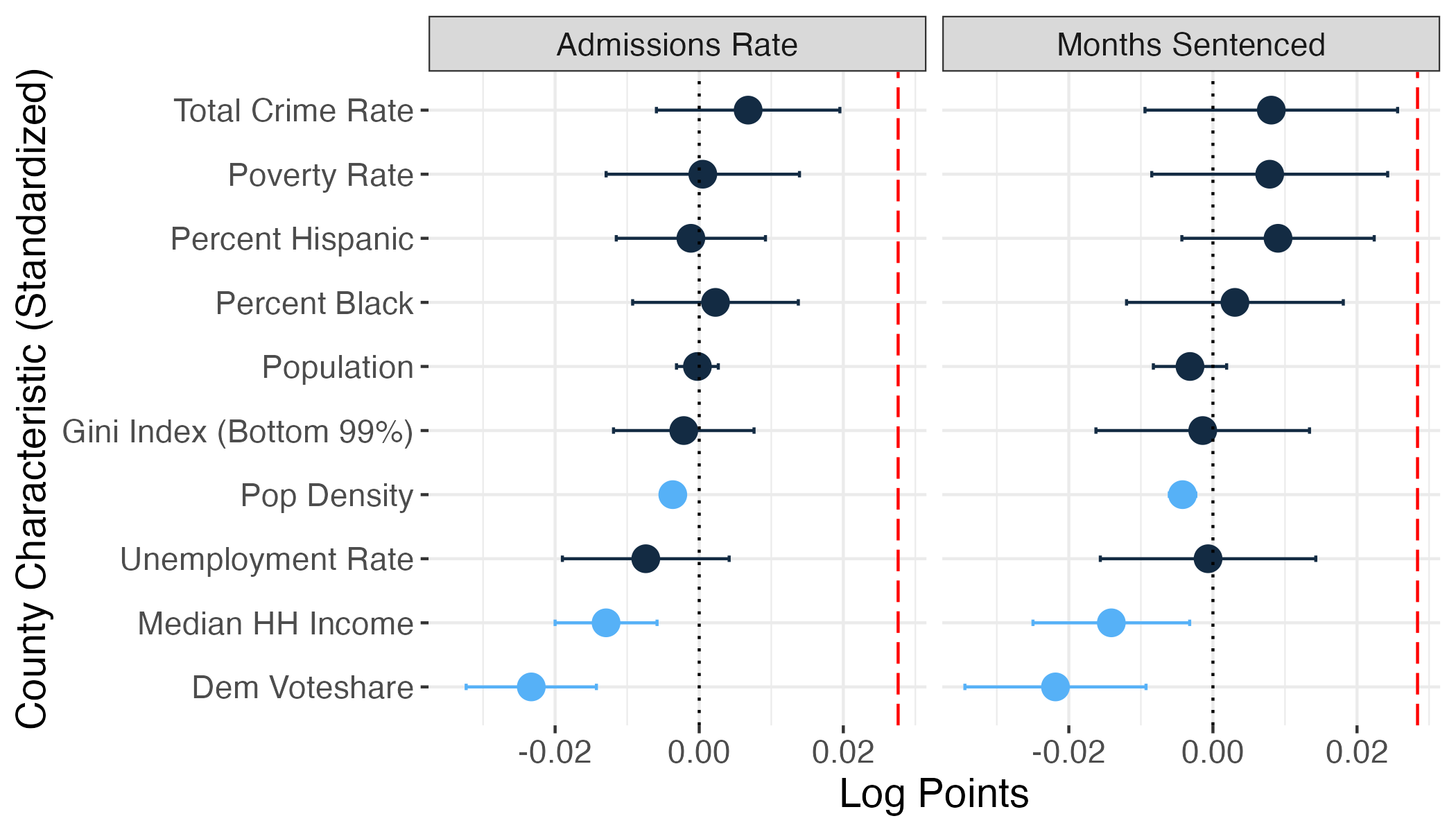} 
\par
\justify \footnotesize \textit{Notes:} The source of county characteristics data is Opportunity Insights (www.opportunityinsights.org).  Estimates reflect a static difference-in-differences model (population weighted---except for the population interaction---to adjust for differences in sampling probabilities across districts and time).  Estimates are at the county level by year.  The standard errors are corrected for correlation across district attorneys and over time in a given county by clustering at the district level.  Estimates include district- and year- fixed effects.  The population interaction is estimated without population weights.  County characteristics are standardized, so estimates reflect the change in election year effects from a one standard deviation change in the corresponding characteristic.  The horizontal lines mark two standard errors.  The vertical dashed line represents the main election year effect.  Dem Voteshare is the vote share in the previous Presidential election going to the Democratic candidate.  All county characteristics are calculated from the mean value between 1986--2006.  
\label{fig:countycharacteristics}
\end{figure}

To deepen our understanding of potential mechanisms explaining this relationship between public sentiment and election year effects, I calculate the association between various standardized county characteristics and election year effects.  Figure \ref{fig:countycharacteristics} shows the result of this analysis, based on the static difference-in-differences model (Equation \ref{eq:staticdid}).  The red vertical line represents the main effect---the election year effect in the absence of interactions with county characteristics.  Figure \ref{fig:countycharacteristics} shows that democratic voteshare, which measures the fraction of votes going to the Democratic Presidential candidate in the prior election, is the most influential county characteristic in the plot.  It is negatively associated with the election year effects, and it is statistically significant for both outcomes.  Again, we see that the political ideology of the district is associated with election year effects, which is consistent with the notion that it is public opinion that is creating political incentives for district attorneys. \par

\subsection{DA Election Effects Track Public Punitiveness: Larger Across Tough-on-Crime Era; Lower After}

\begin{figure}[!tb]
\captionsetup{justification=centering}
\caption{Election Year Effects Over Time vs.\\ Public Opinion Over Time}
\centering
\includegraphics[width=1\textwidth]{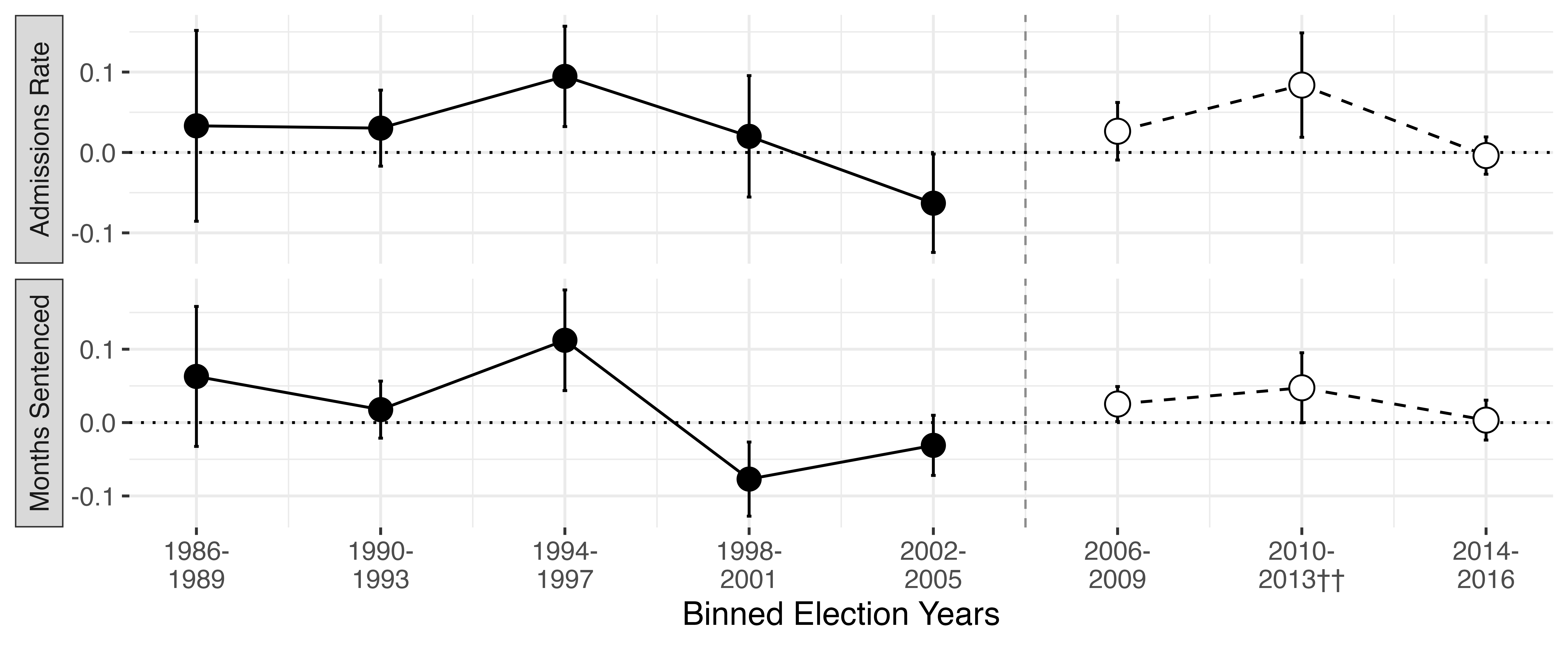} 
\par
\justify \footnotesize \textit{Notes:} The graph depicts a static difference-in-differences model (population weighted to adjust for differences in sampling probabilities across districts and time).  Estimates are at the county level by year.  The standard errors are corrected for correlation across district attorneys and over time in a given county by clustering at the district level.  Estimates include district- and year- fixed effects.  Vertical lines mark two standard errors.  Hollow points (2006--2016) reflect estimates of the period following the era of rising incarceration, which lies outside the main analysis sample and is not restricted to districts with identified prosecutors.  The vertical dashed line marks the boundary.  \\$^{\dagger\dagger}$The 2010--2013 estimates incorporate California's October 2011 Public Safety Realignment (AB 109), which sharply reduced prison admissions in that cycle's comparison years---likely inflating the estimate for the 2010--2013 bin.  Excluding California, the 2010--2013 estimates become not statistically significantly different from zero: -0.003 (s.e. 0.009) for admissions per capita and -0.011 (s.e. 0.009) for sentenced months per capita (Appendix
Table C.17).\par \vspace{.8em}
\centering
\includegraphics[width=1\textwidth]{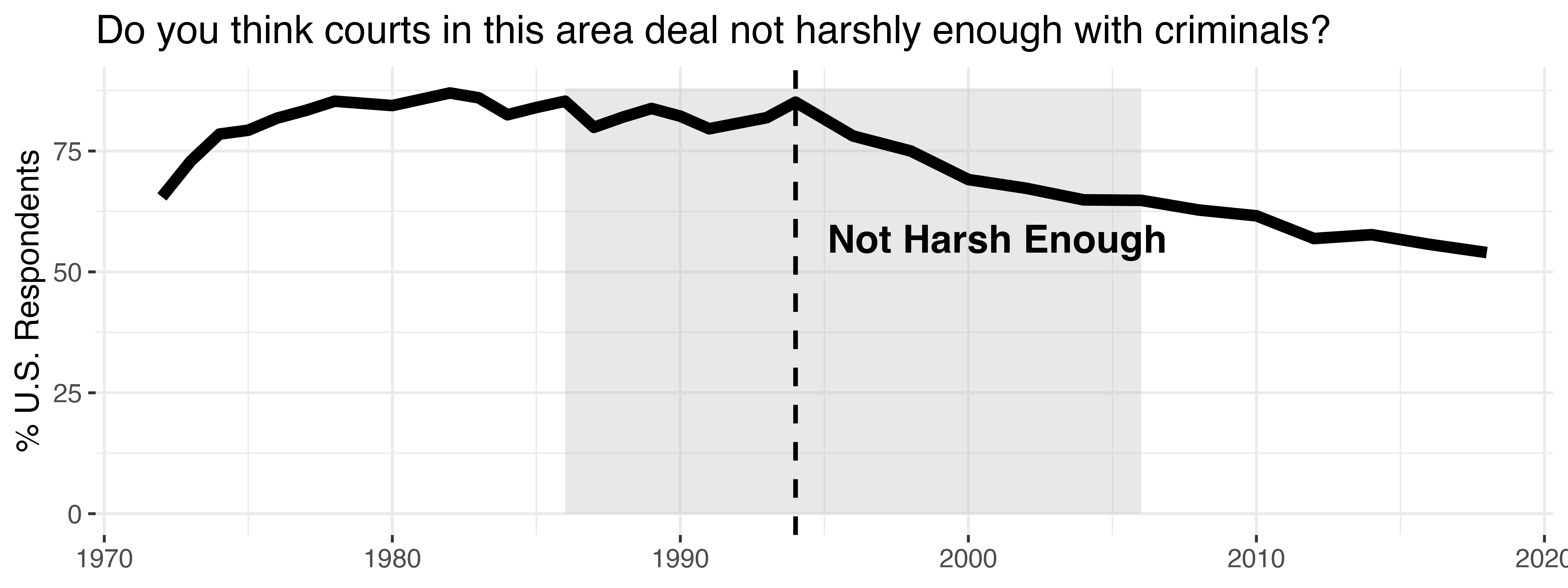}
\justify \footnotesize \textit{Notes:} Source is the GSS \citep{smith2018}. ``Don't Know" responses excluded.  Vertical dashed line corresponds with the year 1994.  Shaded region corresponds with the approximate coverage of core empirical analysis illustrated in Figure \ref{fig:prisonpop} and explained in Section \ref{sec:data} (an era of rising incarceration).
\label{fig:temporalanalysis}
\end{figure}

The change in election year effects over time in the top plot of Figure \ref{fig:temporalanalysis} is consistent with the view that evolution in public sentiment over time impacted DA behavior.  In the 1980s and 1990s, anecdotally there was a lot more discussion about---and perhaps perceived benefit from---being a politician who was viewed as being ``tough-on-crime."  This was the period of the spike in crack-related drug convictions, the passing of the 1994 crime bill, and the establishment of minimum sentencing laws around the country.  The GSS shows that, in fact, the vast majority of the population believed courts were not harsh enough.  The bottom plot of Figure \ref{fig:temporalanalysis} shows that only after 1994 did this fraction of the population begin to decline.  In contrast, those who believed courts were too harsh or about right have become a growing share of the population since 1994.  In short, public sentiment toward punishment has been shifting over the last 25 years or so.  This shift coincides with lower election year effects over time, and plausibly impacted the political incentives faced by district attorneys. \par

The post-2005 period in Figure \ref{fig:temporalanalysis} shows no consistent electoral cycle.  Admissions rate estimates for 2006--2009 and 2014--2016 do not statistically differ from zero, consistent with gradual rather than abrupt attenuation as public punitiveness receded.
Though 2010--2013 does have a positive estimate, this is explained by the enactment of a major state policy (rather than electoral timing)---specifically, the implementation of California's 2011 Public Safety Realignment.\footnote{California's Public Safety Realignment Initiative went into effect in 2011 and was intended to reduce the state prison population by shifting lower-level offenders previously bound for state prison to county jails.  The main dataset for this paper, NCRP, tracks state prison admissions, so this policy shock reduced the incidence of prison admissions captured by NCRP for California beginning in 2011.}  As nearly all California district attorney elections in this bin occurred in 2010, this 2011 policy shock sharply reduced prison admissions across California's non-election years within that bin, mechanically inflating the election year estimate.  Appendix Table C.17 corroborates this point, showing that excluding California makes the 2010--2013 estimate statistically indistinguishable from zero, consistent with the null estimates in the other post-era bins.\footnote{See Appendix Figures C.7--C.8 for more estimates suggesting no election effects persisted following the era of rising incarceration.}

\begin{figure}[!tb]
\captionsetup{justification=centering}
\caption{Salience of Leading Local Newspaper Articles with Crime-Related Language, 1991--2008}
\centering
\includegraphics[width=1\textwidth]{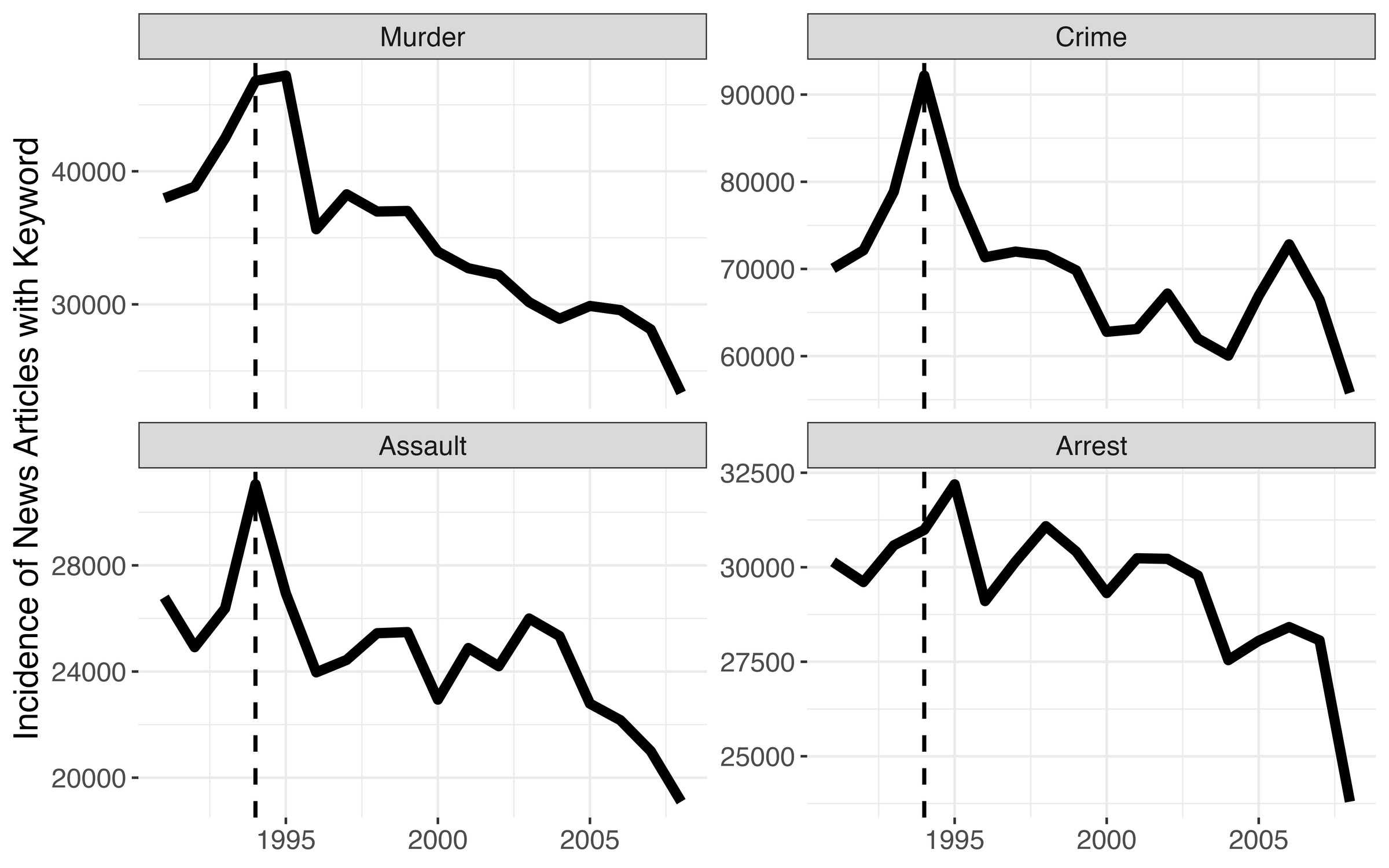} 
\par
\justify \footnotesize \textit{Notes:} Article counts taken from one of the top three newspapers in each respective state.  Searches were performed via Factiva, Nexisuni, Hollis, or WestlawNext.  Given data availability, analysis begins from 1991 onward and includes the following states: Alaska, Arizona, California, Colorado, Connecticut, District of Columbia, Florida, Georgia, Kansas, Kentucky, Louisiana, Maryland, Massachusetts, Minnesota, Missouri, Nevada, New Hampshire, New York, North Carolina, Ohio, Oklahoma, Oregon, South Carolina, Texas, Utah, Virginia, Washington, and Wisconsin.  The newspapers used in the analysis can be found in Appendix Table C.23.

\label{fig:newspaper}
\end{figure}

There is plausibly a nexus between (1) changes in public opinion over time, and (2) changes in the salience of criminal activity over time.  In other words, the public may adopt a more lenient stance toward criminal punishment in periods in which crime is less prominent.  The guiding hypothesis is clear: the \textit{salience} of crime is plausibly more impactful on public sentiment---and in turn the political incentives district attorneys face---than the actual crime rate, since salience more proximately influences voter perceptions than the crime levels themselves.  Under this hypothesis---and based on the public sentiment trends described in Figure \ref{fig:temporalanalysis}---one would expect that the softening of public opinion toward punishment would coincide with (and potentially be caused by) a declining salience of criminal activity in the public eye.  The GSS figure at the bottom of Figure \ref{fig:temporalanalysis} illustrates a clear shift in the public opinion trend line after the year 1994.  Now, compare that shift in trends with Figure \ref{fig:newspaper}.  Figure \ref{fig:newspaper} shows the incidence of top local newspaper articles that include various crime-related keywords; in short, it can be viewed as a proxy for the salience of crime in the public eye.  Strikingly, again we see a clear break in the trend line around the year 1994, after which the incidence of articles mentioning various crime-related language begins to decline.  This general decline occurs for articles that include the words ``murder," ``crime," ``assault," and ``arrest."  Yet, the downward trend is most apparent for articles that contain the word ``murder."  Murders may possess the greatest nexus with public sentiments toward criminal punishment (given the extreme severity and grotesqueness of the offense).  Hence, the sharp decline in local newspaper coverage of murder (and other crime-related terms) after 1994 is consistent with the perspective that the softening of public sentiment toward punishment is in response to the declining salience of criminal activity in the public eye. \par

Appendix Figure C.8 shows monthly estimates for both the admissions rate and the total months sentenced across offense subcategories after 2005 (with adjustments for seasonality performed in an equivalent manner as in Section \ref{sec:monthly}).  Unlike Figure \ref{fig:offensesmonthly} (during the era of rising incarceration), there is either no clear observable increase in the level of criminal sentencing around the time of the election or the increase persists for so many months after the time of the election to be plausibly attributable to DA behavior during election periods.

\section{Conclusion} \label{sec:conclusion}

For this paper, I compiled a new comprehensive dataset on the election cycles and offices sought for all district attorneys in office during the steepest rise in incarceration in U.S. history (roughly 1986--2006).  Using quasi-experimental methods, I found that being in a DA election year increased total admissions per capita to state prisons and total months sentenced per capita.  I found evidence that sentencing outcomes are associated with public sentiment: (1) election effects are higher in the South; (2) election effects depend on county political ideology, being higher in Republican counties; (3) election effects declined in the era of rising incarceration, coinciding with softening public opinion on punishment as measured through the General Social Survey; and (4) election effects disappeared at the national level after the era of rising incarceration ended. \par

If sentencing outcomes were solely determined by the DA’s personal philosophy toward punishment, one might not expect there to be much cyclicality in the severity of outcomes.  One would simply expect the districts with DAs having harsher theories of punishment to have \textit{consistently} more severe sentences.  Instead, we see greater election-year effects in Republican counties, which is consistent with the perspective that DA behavior incorporates political incentives, such as perceived voters' preferences for harshness.  Future analysis might explore ``progressive prosecutors”---based on the finding that political incentives for the level of sentencing differ markedly between liberal and conservative districts, this may translate to different behaviors in the run-up to elections for various DA offices.  Other legal scholars have already touted the potential for progressive prosecutors to help reform the system \citep{hessick2020}.  One would expect not only that areas with progressive prosecutors would have lower overall sentences, but based on this study, one would also expect there to be no election year effects---or perhaps election year effects that are less pronounced---for districts where progressive prosecutors are politically viable. \par

There are at least two obvious interpretations of the election year effects this study finds: (1) that the election year level of sentencing is too high, based on prevailing theories of justice; (2) that the election year level of sentencing is adequate, and instead the level for \textit{non}-election years is too low.  Given the fact that the United States has the highest incarceration rate in the world, many might find the second interpretation less plausible than the first one.  Furthermore, past research has found that initial felony convictions that occurred within six months of district attorney elections had a greater likelihood of being overturned in appellate court, providing evidence that election dynamics may yield more inaccurate sentences, wrongful convictions, and deviations from justice \citep{mccannon2013prosecutor}.  Hence, under the first interpretation that the election year levels of sentencing are too high, politics may have encouraged overly severe sentencing outcomes.  Thus, the argument for reducing the political incentives of DAs becomes more salient---particularly when the public's desire for harshness diverges from what may be considered ``fair" or socially optimal. \par

Accordingly, the findings relate to broader legal debates on the efficacy of ``democratizing” the criminal justice system \citep[see, e.g.,][]{rappaport2020}.  On one side of the argument are those who attribute many problems with the criminal legal system to an ill-informed (or particularly vengeful) American public; this side tends to call for the justice system to be more greatly influenced by professionals and experts.  On the other side of the argument are those who attribute the problems with the criminal legal system to bureaucratic attitudes and other institutional/structural problems.  This side tends to call for more control of the system from local communities, with greater responsiveness to the lay public.  For those skeptical of democratizing the criminal legal system, a potential corrective response to the election year effects could be to transition the DA selection process from an electoral system to an appointment one.  Several states/areas already operate under this model, such as New Jersey, Alaska, Connecticut, and Washington, DC. \par

Some legal scholars, however, question the political feasibility of accomplishing this shift \citep{kress1976, wright2014}.  For those who support increased democratization of the criminal legal system, appropriate corrective responses to the election year effects found in this paper are less clear.  According to some, even if criminal justice actors were made more accountable to local communities, it remains unclear whether the lay public in the local communities most afflicted by crime would necessarily encourage more leniency \citep{rappaport2020}.  There could be unintended consequences, with more democratization potentially yielding political incentives to \textit{increase} the severity of sentences.  Similarly, proposals to increase the competitiveness of DA elections may exacerbate the impact of election cycles on sentencing outcomes \citep{hessick2020}.  Ultimately, the myriad consequences of further democratization in the DA selection process may remain an empirical question---one that I leave for future research. \par

Of note, this paper only evaluates the short-term cyclical impact of being in an election year on sentencing outcomes.  It does not capture potential longer-term political considerations that might impact how district attorneys approach their work at \textit{all} periods in the election cycle.  This means the full impact of political incentives on the criminal legal system and the unprecedented rise in incarceration may be larger than what is found in this paper alone: not only (1) might DAs be responding to political incentives in ways above and beyond what can be captured through this paper's empirical strategy alone, but also (2) the findings may signal that \textit{other} elected officials, particularly legislators, were also responding to public sentiment in making legislative choices during the era of rising incarceration.  So much work on criminal justice reform to date has focused on public policy improvements to the system itself and to the laws that govern it.  If the political incentives of those entrusted to administer justice are shaped by public sentiment regarding criminal punishment, then this paper carries two implications.  First, for political economy: electoral accountability in the criminal legal system may operate even absent competitive elections, through career concerns and anticipatory behavior that respond to voter preferences. Second, for public policy: focusing on ``hearts and minds"---on shifting public opinion toward punishment independent of shifts to laws or to the structure of the criminal legal system itself---may prove not only beneficial but also instrumental in improving sentencing outcomes and stemming historic levels of incarceration.

\pagebreak

\section*{Acknowledgments} \label{sec:conclusion}

I extend deep thanks to Isaiah Andrews, Raj Chetty, Nathan Hendren, Lawrence Katz, Michelle Pearse, Maya Sen, James Snyder, and---above all---to God.  I am greatly appreciative of thoughtful and very constructive comments from Co-Editor Thomas Fujiwara and two anonymous referees. I also thank Daron Acemoglu, Alberto Alesina, Ian Ayres, Keith Chen, Melissa Dell, Angela Dixon, David Ellwood, William Hubbard, Louis Kaplow, Yair Listokin, Jonathan Masur, Nathan Nunn, Ben Olken, John Rappaport, Ben Schneer, Theda Skocpol, Elie Tamer, Crystal Yang, and participants in the Law and Economics Seminar (Harvard Law School), Law and Economics Workshop (UC Berkeley Law), Law and Economics Workshop (Georgetown Law School), and Work-In-Progress Seminar (University of Chicago Law School).  I also give thanks for outstanding research assistance from Montse Trujillo, My Vuong, Amy Quan, Ramiz Razzak, and Matej Cerman.  

Funding: This work was supported by the Association for Social Economics, the Eisenhower Institute at Gettysburg College, the Ford Foundation Predoctoral Fellowship, the Foundations of Human Behavior Initiative, the Harvard Institute for Quantitative Social Sciences, the Harvard Lab for Economic Applications and Policy, the Harvard Multidisciplinary Program in Inequality \& Social Policy, the Horowitz Foundation for Social Policy, the Radcliffe Institute for Advanced Study, and the National Institute on Aging, Grant Number T32-AG000186.

\pagebreak

\printbibliography

\clearpage

\renewcommand{\thefigure}{\arabic{figure}}

\end{document}